\documentclass[manuscript]{acmart}

\AtBeginDocument{%
  }

\setcopyright{acmlicensed}
\copyrightyear{2024}
\acmYear{2024}
\acmDOI{XXXXXXX.XXXXXXX}

\acmConference[Conference acronym 'XX]{Make sure to enter the correct
  conference title from your rights confirmation emai}{June 03--05,
  2018}{Woodstock, NY}

\acmISBN{978-1-4503-XXXX-X/18/06}

\usepackage{geometry}
\usepackage{array}
\usepackage{longtable}
\usepackage{multirow}
\usepackage{array} 
\usepackage{float}

\begin{document}

\title{KoroT-3E: A Personalized Musical Mnemonics Tool for Enhancing Memory Retention of Complex Computer Science Concepts.}

\author{Xiangzhe Yuan}
\affiliation{%
  \institution{City University of Hong Kong}
  \city{Hong Kong}
  \country{China}}
\email{yuanxiangzhe9@gmail.com}

\author{Jiajun Wang}
\affiliation{%
  \institution{City University of Hong Kong}
  \city{Hong Kong}
  \country{China}}
\email{jiajwang02@gmail.com}

\author{Siying Hu}
\affiliation{%
  \institution{City University of Hong Kong}
  \city{Hong Kong}
  \country{China}}
\email{siyinghu-c@my.cityu.edu.hk}

\author{Andrew Cheung}
\affiliation{%
  \institution{Independent Researcher}
  \city{Hong Kong}
  \country{China}}
\email{andrew.chu.cheung@gmail.com}

\author{Zhicong Lu}
\affiliation{%
  \institution{City University of Hong Kong}
  \city{Hong Kong}
  \country{China}}
\email{zhiconlu@cityu.edu.hk}

\renewcommand{\shortauthors}{Yuan et al.}

\begin{abstract}
As the demand for computer science (CS) skills grows, mastering foundational concepts is crucial yet challenging for novice learners. To address this challenge, we present KoroT-3E, an AI-based system that creates personalized musical mnemonics to enhance both memory retention and understanding of concepts in CS. KoroT-3E enables users to transform complex concepts into memorable lyrics and compose melodies that suit their musical preferences. We conducted semi-structured interviews (n=12) to investigate why novice learners find it challenging to memorize and understand CS concepts. The findings, combined with constructivist learning theory, established our initial design, which was then refined following consultations with CS education experts. An empirical experiment(n=36) showed that those using KoroT-3E (n=18) significantly outperformed the control group (n=18), with improved memory efficiency, increased motivation, and a positive learning experience. These findings demonstrate the effectiveness of integrating multimodal generative AI into CS education to create personalized and interactive learning experiences.
\end{abstract}

\ccsdesc[500]{Applied computing~Interactive learning environments}
\ccsdesc[500]{Human-centered computing~Interactive systems and tools}

\keywords{Music, Mnemonics, Generative AI, Education, Learning, Large Language Models, Computer Science}

\begin{teaserfigure}
  \includegraphics[width=\textwidth]{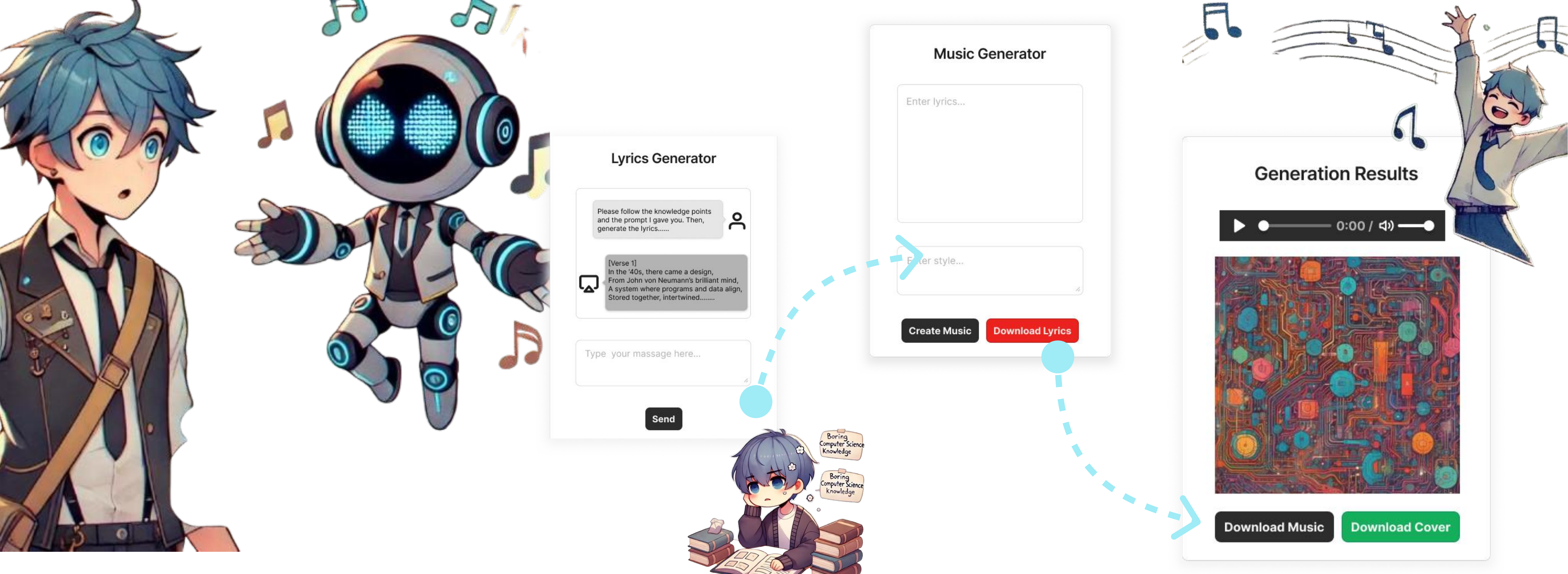}
  \caption{An example workflow of interacting with KoroT-3E: Start with the lyrics generator by entering the concepts to be adapted and providing the relevant prompts; receive and copy the adapted lyrics or download them as a txt file; paste the lyrics into the music generator and specify your preferred music style; obtain the generated musical mnemonic along with a customized music cover, which can be downloaded as well.}
  \Description{Manga B-Roll}
  \label{fig:teaser}
\end{teaserfigure}

\received{20 February 2007}
\received[revised]{12 March 2009}
\received[accepted]{5 June 2009}

\maketitle
\section{INTRODUCTION}
As the information age continues to advance. computers play an increasingly important role in people's lives, and the demand for computer science skills across various industries has significantly increased \footnote{https://www.technologyreview.com/2023/04/20/1071291/learn-to-code-legacy-new-projects-education/}. More students are choosing computer science(CS) as their major, and many non-computer science professionals are also learning relevant CS technologies to enhance their competitiveness\footnote{https://m.163.com/dy/article\_v2/GD6CB9960511RQUG.html}. To support learners in CS education, numerous tools have emerged, but most focus primarily on assisting programming and programming education\cite{winkler2020sara,arawjo2017teaching,kazemitabaar2023studying,kelleher2007storytelling,tang2024vizgroup}, with few supporting the learning of fundamental concepts and underlying CS logic. CS education is not just about coding; it involves understanding the principles of computers, hardware and software design, implementation, and their impact on society\cite{10.1145/2593247,shackelford1997introducing}. Mastering fundamental CS concepts and theories is more crucial than simply learning application tools and programming skills, which can rapidly become outdated\cite{romeike2019role}. A solid understanding of these core principles enables students to better navigate and adapt to the constantly evolving landscape of technologies and applications, fostering sustainable skills for long-term growth\cite{romeike2019role,dickey2023innovating,belowguide}. 

For CS novice learners, one of the biggest challenges is the vast and abstract nature of CS concepts, logic, and formulas\footnote{https://www.jamiefosterscience.com/is-computer-science-hard-for-someone-with-no-experience/}. Many of these concepts are too complex and challenging for novice learners to memorize and understand\cite{soll2022were}. These issues increase novice learners' learning difficulty and affect their motivation and enthusiasm\cite{soll2022were}. Memory-aid tools could potentially be useful in addressing these challenges. For example, the ancient Greeks invented a set of memory-improving techniques called \textit{mnemonics}, which are now widely used in education\cite{yates1966art,putnam2015mnemonics}. Mnemonics can be powerful learning tools in certain contexts\cite{bower1970analysis} because they can enhance memory by utilizing natural memory processes, making them effective for recalling specific types of information\cite{putnam2015mnemonics}. In CS education, previous research has adapted mnemonics to assist students in memorizing SQL concepts and has shown satisfying outcomes\cite{balaji2013mnemonics}.

As classic mnemonic methods,\textit{musical mnemonics} use melodies and rhythms to enhance memory. Musical mnemonics can significantly improve students' memory and recall abilities compared to traditional verbal mnemonics\cite{cirigliano2013musical,joanna2010hot}. Musical mnemonics have already proven effective in subjects like biology\cite{yeoh2014musical} and medicine\cite{koksal2013impact}, which also involve numerous complex and abstract concepts like CS. Moreover, many studies have shown that music can be beneficial in learning and education, such as helping with memory\cite{groussard2010music}, improving students' self-confidence and concentration\cite{cools2023effects}, and cultivating their interest and enthusiasm for learning\cite{yang2023impact}.

However, writing suitable mnemonics usually requires the assistance of experienced individuals or even experts\cite{putnam2015mnemonics}. This poses a challenge for novice learners. Traditional musical mnemonics are typically standardized and pre-written by teachers, but they may lack personalization and fail to meet the diverse needs of all learners. According to the \textit{constructivist learning theory}, when learners actively construct their understanding of knowledge through their own experiences and practices, it is more conducive to developing their independent learning abilities and overall competence\cite{tobias2009constructivist}. Moreover, the \textit{generation effect} suggests that learners generally achieve better memory retention when they create their own study materials rather than relying on those provided by the teacher\cite{bertsch2007generation}. McCabe et al.\cite{mccabe2015learning} demonstrated that the generation effect also applies to mnemonics. It is essential for learners to actively create mnemonics based on their background, experience, and cognitive style.

In this study, Based on the findings from the formative study, the constructivist learning theory, and the consultation of CS education experts, we designed a hybrid AI system, KoroT-3E, to generate musical mnemonics for CS novice learners. Our system (\autoref{korot}) consists of three components: \textit{lyrics generation},\textit{music generation}, and \textit{music display}. For the lyrics generation component, we employed GPT-4o\footnote{https://openai.com/index/chatgpt/} with prompt engineering, while for the music generation component, we utilized Suno\footnote{https://suno.com/}. Users can customize and adapt CS concepts into easy-to-remember lyrics according to their musical preferences and create a suitable melody to form musical mnemonics.

We conducted an experiment to evaluate the effectiveness of KoroT-3E in supporting short-term and long-term memory retention for different types of knowledge. The results of the three tests based on two CS concepts showed that the overall performance of the experimental group (n=18) was significantly better than that of the control group (n=18) (p < 0.05). The experimental group's average scores were consistently higher than those of the control group in each individual test, although only one of these results reached statistical significance. Subsequent surveys and interviews revealed that participants found KoroT-3E easy to use and felt it enhanced their memory retention, improved memory efficiency, and increased their interest and motivation in learning. They also believed that KoroT-3E could be applied to a broader range of fields and user groups. 

The system provides a learning and memorization aid for CS novice learners by adapting concepts they find difficult to remember and understand into personalized musical mnemonics. It demonstrates the potential of generative AI-based mnemonic techniques, particularly in improving the learning of foundational concepts in computer science. Marking a significant step toward developing interactive, personalized learning experiences. Moreover, this system offers a novel approach to modernizing and advancing mnemonic strategies.

This research makes the following contributions to HCI:
\begin{itemize}
 \item We developed a personalized musical mnemonic learning tool using multimodal generative AI, introducing an advanced method that refines mnemonic techniques and offers a novel approach to learning foundational concepts in computer education.
 
 \item Empirical results from the user evaluation (n=36), which demonstrated the system's effectiveness in enhancing both short-term and long-term memory retention across different types of knowledge.

\end{itemize}

\section{RELATED WORK}

The existing body of research covers the use of mnemonics and music in education and the application of advanced technologies in HCI for CS education. To frame our investigation, we first analyze the diverse roles of mnemonics and music in enhancing learning across different educational contexts. We then explore how emerging technologies, such as VR, AR, and AI, have been leveraged in HCI to support CS education. However, we identify a gap in current research: limited focus on supporting learning foundational CS concepts. Our work aims to fill this gap by developing innovative tools to enhance learning outcomes for CS concepts.

\subsection{Mnemonics}
Mnemonics are memory enhancement strategies that facilitate the recall of information by transforming it into more memorable forms. The core idea behind mnemonics is to leverage existing memories or associations, integrating new information into an established cognitive framework to improve memory retention\cite{akpan2021impact,roediger1980effectiveness,mccabe2013psychology}. Recent research has further discovered that mnemonic training can alter the brain's functional network structure, thereby supporting enhanced memory capabilities. This provides new insights into brain plasticity\cite{dresler2017mnemonic}. The use of mnemonics dates back to ancient Greece, with the rhetorician Simonides often credited as a pioneer in this field for developing the "Method of Loci."\cite{akpan2021impact}. Mnemonics can be categorized into various types based on their characteristics and applications, including keyword mnemonics, imagery mnemonics, acronyms, and musical mnemonics. Keyword mnemonics enhance memory by associating new information with familiar keywords through phonetic or semantic connections \cite{king1992toward}. Imagery mnemonics utilize vivid images or visual cues to aid in retaining complex information\cite{kaschel2002imagery,richardson1995efficacy}. Acronyms combine the initial letters of a series of words into a memorable phrase or word\cite{lewis2018importance}. Musical mnemonics, on the other hand, involve setting information to melody and lyrics, which enhances both memory retention and recall \cite{cirigliano2013musical}. Mnemonics have been widely applied across various fields beyond their most common educational use. They are also utilized in healthcare, such as aiding individuals with memory impairments\cite{kaschel2002imagery} or Alzheimer's disease\cite{sahakian1990sparing}, and in areas like security and privacy\cite{woo2016improving,kuo2006human}, commercial advertising\cite{yalch1991memory,smith2001age} et al. These studies suggest that mnemonics effectively enhance memory across different domains, indicating a promising potential for broader applications.

\subsubsection{Mnemonics in Education}

In the field of education, mnemonics are widely used to enhance learning and memory, with studies showing their significant effectiveness across various subjects and age groups for different learning tasks\cite{raugh1975mnemonic,manalo2004using}. For instance, in early literacy instruction for kindergarten students, mnemonics have been shown to significantly improve letter recognition and phonetic skills\cite{agramonte2002using}. Mnemonics are also effective for college students. An experiment by Kaur et al.\cite{kaur2022effect} involving 97 third-year undergraduate nursing students demonstrated that a teaching approach incorporating mnemonics significantly improved memory retention and learning outcomes compared to traditional lecture methods alone. Mnemonics are often perceived as more suitable for fields or knowledge areas that require extensive memorization, such as language learning\cite{azmi2016case}, rather than for mastering higher-order skills like comprehension or knowledge transfer\cite{putnam2015mnemonics}. However, some studies have shown that mnemonics can be effective in more complex learning areas as well. In mathematics, research indicates that mnemonics can help elementary students better grasp mathematical concepts, particularly when memorizing complex concepts or steps\cite{delashmutt2007study}. In biology, research by Yeoh demonstrated that musical mnemonics significantly enhanced memory retention and recall of complex biological processes, such as RNA Transcription\cite{yeoh2014musical}, Protein Synthesis\cite{yeoh2014musical1}, and the Calvin Cycle\cite{yeoh2013musical}. 

Additionally, there is limited evidence suggesting that mnemonics can directly support higher-order learning. For example, Carney et al.\cite{carney2003promoting} found improved performance on tests involving hierarchical relationship reasoning. However, more research is needed to substantiate these findings. While mnemonics have proven effective in many areas, they also have certain limitations. For example, their applicability may not extend to all domains; they depend on individual preferences and require learning time\cite{putnam2015mnemonics}. Although many regard mnemonics as a revolutionary learning tool, they are not a panacea in educational practice\cite{levin1993mnemonic}. Mnemonics are unlikely to completely transform educational systems, but under appropriate conditions, they can still serve as an effective tool for learning\cite{putnam2015mnemonics}.

\subsubsection{Mnemonics in HCI}
Research on mnemonics within the field of HCI remains relatively limited. J{{\"A}ngeslev{\"a} et al.\cite{angesleva2003body} explored using the user's physical space as an interactive interface, leveraging embodied cognition to reduce cognitive load while employing motion sensing to facilitate information storage and retrieval. Ikei et al.\cite{ikei2007spatial} enhanced traditional mnemonic techniques with electronic devices by digitally annotating objects, locations, and people in real spaces, creating an "external virtual memory space (eVMS)" to improve memory performance. Perrault et al.\cite{perrault2015physical} utilized spatial, object, and semantic memory to enhance users' ability to select and remember gesture commands in smart home environments.

\subsection{The Benefits of Music in Education}

In the field of education, the application of music has increasingly drawn attention. Numerous studies have explored the various uses and impacts of music in different educational settings, ranging from traditional classroom environments\cite{campabello2002music} to interdisciplinary education\cite{scripp2002overview}. These studies cover a wide spectrum, from cognitive improvement to emotional and social development\cite{hallam2016impact,yang2023impact}. Research indicates that students who participate in music education tend to perform better across multiple subjects, particularly in English and mathematics, than those who do not \cite{hallam2016impact}. Similarly, another study found that music education positively impacts students' academic motivation, especially by enhancing their enjoyment, sense of competence, and interest in learning \cite{cools2023effects}. These findings suggest that music education not only helps improve cognitive abilities but also fosters overall academic performance by increasing student engagement and enthusiasm\cite{scripp2002overview}.

Additionally, music has been used in both formal and informal educational settings to promote social-emotional learning. Research found that music can enhance student's learning and well-being, revealing that music can help reduce stress, improve mood, strengthen social relationships, and enhance overall well-being \cite{hu2021university}. 
The educational benefits of music are also evident in its support for creativity and cultural diversity. Qi \cite{qi2023role} highlights that through learning and creating music, students are exposed to music from diverse cultural backgrounds, which stimulates their creative thinking. 
In summary, music is an engaging and effective approach to enhancing children's cognitive functions\cite{tervaniemi2018promises}, contributing to increased cognitive reserve and thereby slowing age-related memory decline\cite{groussard2010music}. Applying music in the educational field proves to be highly effective.

\subsection{CS Concepts Education in HCI}
In HCI, the rise of emerging technologies has introduced innovative tools and techniques to enhance CS education. For example, a 3D animation-based programming environment developed by Kelleher et al. \cite{kelleher2007storytelling}. uses storytelling to make programming more engaging, significantly increasing interest and participation in programming. Wang et al.\cite{wang2012block} invented E-Block leverages wireless and infrared technologies to support children's understanding of fundamental programming concepts through physical modules and maze games. Wang et al.\cite{wang2021puzzleme} use Puzzleme to improve learning outcomes in programming by facilitating peer support in the classroom without increasing teaching costs.

However, these tools primarily focus on programming, while there are relatively few learning aids aimed at CS Concepts. SmileyCluster, created by Wan et al.\cite{wan2020smileycluster}, for instance, developed a learning environment based on the “face overlay” conceptual metaphor, using visualization techniques and interactive exploration to enhance students' understanding of machine learning methods. Shastri et al.\cite{shastri2020machine} different initiative introduced a machine learning course for non-programmers, using the open-source data mining tool Orange to help learners construct and evaluate ML models. Consequently, developing more tools that integrate these advanced technologies to assist in understanding core CS concepts presents a promising opportunity.

\begin{figure}[t]
    \centering
    \includegraphics[width=\linewidth]{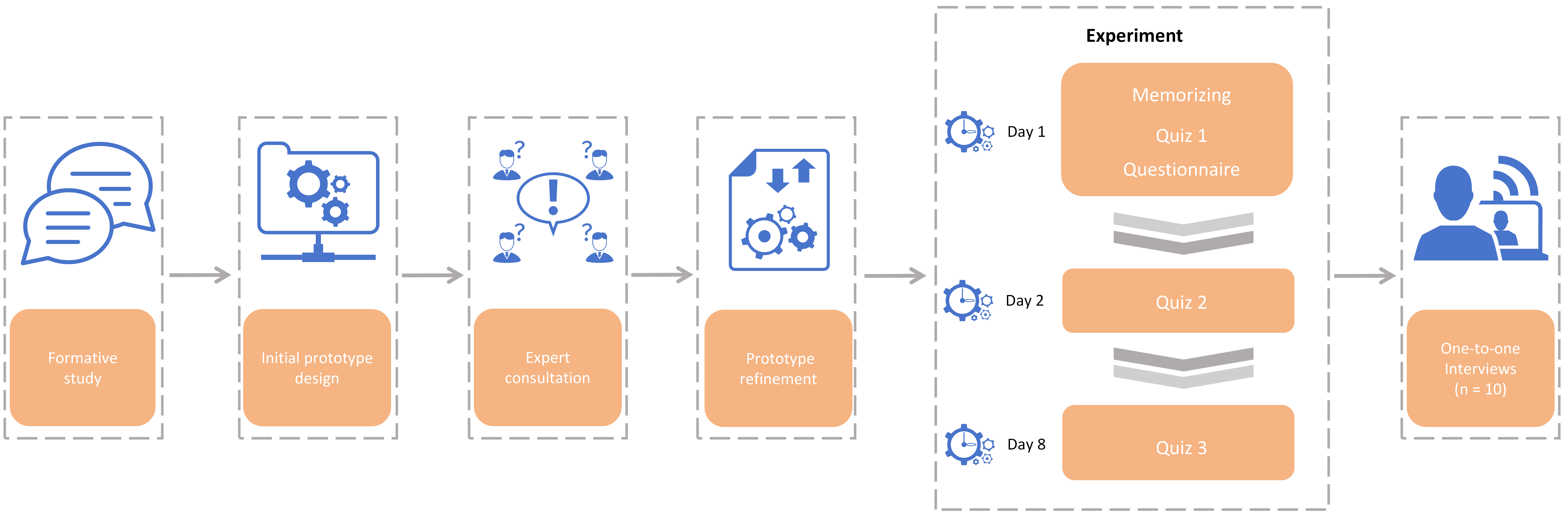}
    \caption{Overall research design and procedure.}
    \label{procedure}
\end{figure}

\section{FORMATIVE STUDY}

To understand the challenges CS novice learners encounter in remembering and comprehending fundamental concepts while learning computer science. We interviewed 12 participants, including seven individuals with formal backgrounds in computer science and five self-taught individuals currently engaged in computer-related work and research. All participants have approximately four years of learning experience. The rationale for selecting individuals with around four years of experience is twofold. First, the rapid evolution of the computer science field means that insights from those with significantly more experience may not apply to current learning practices, as much of what they learned may no longer be mainstream. Second, four years is a typical duration for completing an undergraduate degree in computer science, ensuring that these individuals have recently completed systematic studies and possess experience across various field areas. Understanding the perspectives of self-taught professionals is also crucial, given the growing number of individuals entering the field through non-traditional paths. Chen et al.\cite{chen2024learning}found that non-traditional programmer groups, who typically lack a CS background and have not undergone formal programming education, often have a limited understanding of certain programming concepts. They suggest that future research should focus on the needs of non-traditional programming groups, designing tools and resources that better support them. By including these self-learners in our interviews, we aim to capture a comprehensive view of the challenges and experiences of formally educated and self-taught computer science practitioners.

The interviews were conducted in a semi-structured format, primarily focusing on two questions: 1. When learning the basics of computer science, do they encounter any difficult concepts to memorize or understand? 2. If so, what are those concepts? 3. Why do they think these concepts are challenging to remember or comprehend? The interview results are in \autoref{tab:formative_results}.

\subsection{Data Collection and Analysis}

All interview data, including screen captures and audio recordings, were entirely preserved. We manually transcribed the content from the screen captures and audio recordings and conducted thematic analysis in the following manner:
 1. Two researchers independently reviewed and openly coded the records, then met to discuss and inductively generate initial codes. 2. One researcher applied these codes across all interview data, ensuring comprehensive coverage. Any code modifications were discussed between the researchers until consensus was reached. 3. Following initial coding, they collaborated to merge codes into categories, establishing connections and developing themes relevant to the research questions. We conducted the qualitative analysis in Chinese. Finally, we translated the interview quotes into English with the assistance of experienced translation experts and reported our research findings accordingly.

\begin{table*}[t]
\centering
\caption{The results from the interviews conducted during the formative study.}
\begin{tabular}{m{2.5cm}|m{4cm}|m{7.5cm}}
\hline
\textbf{Category} & \textbf{Challenge} & \textbf{Description} \\
\hline
\multirow{3}{*}{\centering CS background} & Complexity and isolation of concepts & Computer science encompasses numerous subfields with complex content, and many of the concepts lack interrelation. \\
\cline{2-3}
& Hard to understand & Much of the knowledge in computer science is abstract and difficult to understand. \\
\cline{2-3}
\hline
\multirow{2}{*}{\centering \begin{tabular}[c]{@{}c@{}}Self-taught\\Professionals\end{tabular}} & Lack of Interest & Since they usually study computer science to complete certain tasks or work purposes, many are not very interested in learning computer science. \\
\cline{2-3}
& Hard to Learn & Many parts are hard to learn through self-study or require a significant amount of time and effort. \\
\hline
\end{tabular}
\label{tab:formative_results}
\end{table*}

\subsection{Challenges of Participants \textit{\textbf{with}} a CS Background}

Based on the interview results, we found that only one participant with a CS background considered learning to code the most challenging aspect. The remaining seven participants believed that understanding the background knowledge of CS was the most difficult. The perspectives of participants with a CS background mainly revolved around three themes: 1) Complexity and isolation of concepts; 2) Too abstract and difficult to understand.

\subsection{Challenges of Participants \textit{\textbf{without}} a CS Background}

Participants with a background in CS and those without a CS background have somewhat different perspectives. Due to the lack of systematic study, their difficulties are relatively more general. Through our coding, we found that their views can mainly be categorized into two themes: 1) Lack of interest. 2) Too much content and difficulty.

\subsection{Conclusion of Formative Study}
Therefore, summarizing all the interviews, we found that the responses of participants with and without a CS background differ in certain ways. Due to the lack of systematic study, participants without a CS background tend to identify challenges at a more macro level and often provide vague reasons. In contrast, participants with a CS background can specify particular courses or fields and provide detailed analyses of why these aspects pose challenges during their initial learning stages. We have synthesized all the challenges mentioned by both groups into three main points: 1)There is an overwhelming amount of content, with many concepts lacking interrelation, making it difficult to remember. 2)The content is too abstract and hard to understand. 3)There is a lack of interest in learning. 

\section{SYSTEM: KOROT-3E}

After synthesizing the feedback from the formative study, we begin our system design. Our design of KoroT-3E is primarily grounded in the Constructivist Learning Theory \cite{hein1991constructivist}. According to this theory, learning is an active process of constructing knowledge rather than passively receiving it\cite{tobias2009constructivist,hein1991constructivist}. Learners actively build their understanding and meaning of knowledge through interactions with their environment, others, and their existing knowledge\cite{bada2015constructivism}. Therefore, our \textbf{first design consideration} emphasizes that the system should allow learners to actively participate in creating musical mnemonics, providing them with personalized musical mnemonics that align with their individual preferences. This personalized learning experience enhances students' motivation and engagement, encouraging them to actively participate in their learning rather than passively absorbing information. The \textbf{second design consideration}, inspired by the social aspect of constructivism, recognizes that social interaction with teachers and peers helps deepen students' understanding of knowledge through sharing and discussion\cite{bada2015constructivism}. To facilitate this, KoroT-3E includes a feature that allows students to save and share their generated lyrics and music, fostering a collaborative learning environment where they can exchange their musical mnemonics with others. As a cultural and emotional context, music provides a realistic and relevant background for learning\cite{campabello2002music} as our \textbf{third design consideration}. By integrating concepts into music, students memorize the information and associate it with the musical context, making it easier to recall and apply the knowledge in future situations. In terms of the \textbf{fourth design consideration}, creating lyrics and generating music encourages students to reflect on their understanding of the material, allowing them to revise and optimize their work. This reflective activity strengthens learning outcomes and helps students learn from and improve upon their mistakes. KoroT-3E supports this by allowing students to repeatedly generate musical mnemonics until they are satisfied with the result.

Since KoroT-3E is a tool designed to aid learning, we believe its interface should adhere to Usability Principles to ensure ease of use and prevent users from spending excessive time learning it. The entire interface focuses on aesthetic and minimalist design to create a simple layout that avoids overwhelming users with unnecessary information. Each input field and button is annotated with its function to ensure that users can quickly understand how to use the tool. \autoref{korot} shows the KoroT-3E interface and usage process. 

To enhance KoroT-3E and better support CS novice learners, we consulted with three CS education experts, each with over five years of experience, during its initial testing and improvement phases. These experts assessed the system's functionality and design from both technical and professional perspectives, providing targeted feedback to refine the concept and user settings. The process involved three steps: presenting the system's development purpose based on prior research, demonstrating its workflow, and seeking expert evaluations on its usefulness, design rationality, and areas needing improvement. All interview data were analyzed as outlined in Section 3.1. Based on expert feedback, we refined the KoroT-3E concept and settings, focusing on the accuracy and reliability of adapted lyrics. To achieve this, we integrated Prompt Engineering, emphasizing clear, precise, and concise prompts to improve response quality \cite{marvin2023prompt, giray2023prompt}. Providing contextual information, such as specific CS concepts, is essential to avoid generic or inaccurate responses \cite{marvin2023prompt}. We recommend KoroT-3E users follow these guidelines to enhance adaptation accuracy. Additionally, after testing various large language models, experts identified GPT-4o as the best performer, which we implemented in KoroT-3E. The entire workflow of the refined KoroT-3E is as follows:

\begin{enumerate}
 \item Lyrics Generation Module: Users use this module to adapt concepts into suitable lyrics. They can iteratively modify the generated lyrics through prompt interactions according to their preferences, making the content more aligned with their expectations and more engaging.

 \item Music Generation Module: Once users are satisfied with the adapted lyrics, they input them into the music generation module. Here, users can also adjust the musical style of the generated musical mnemonics. Additionally, the generated lyrics can be saved for future review.

 \item Music and Image Generation: After finalizing the lyrics and musical style, the system generates the music and returns it to the user, along with a cover image created by Suno based on the lyrics. Users can then play the music and save the music and the cover image for later use.

\end{enumerate}
This design ensures that users are actively involved in the learning process, tailoring the mnemonic aids to their tastes and making the learning experience more enjoyable and effective.

\begin{figure}[t]
    \centering
    \includegraphics[width=\linewidth]{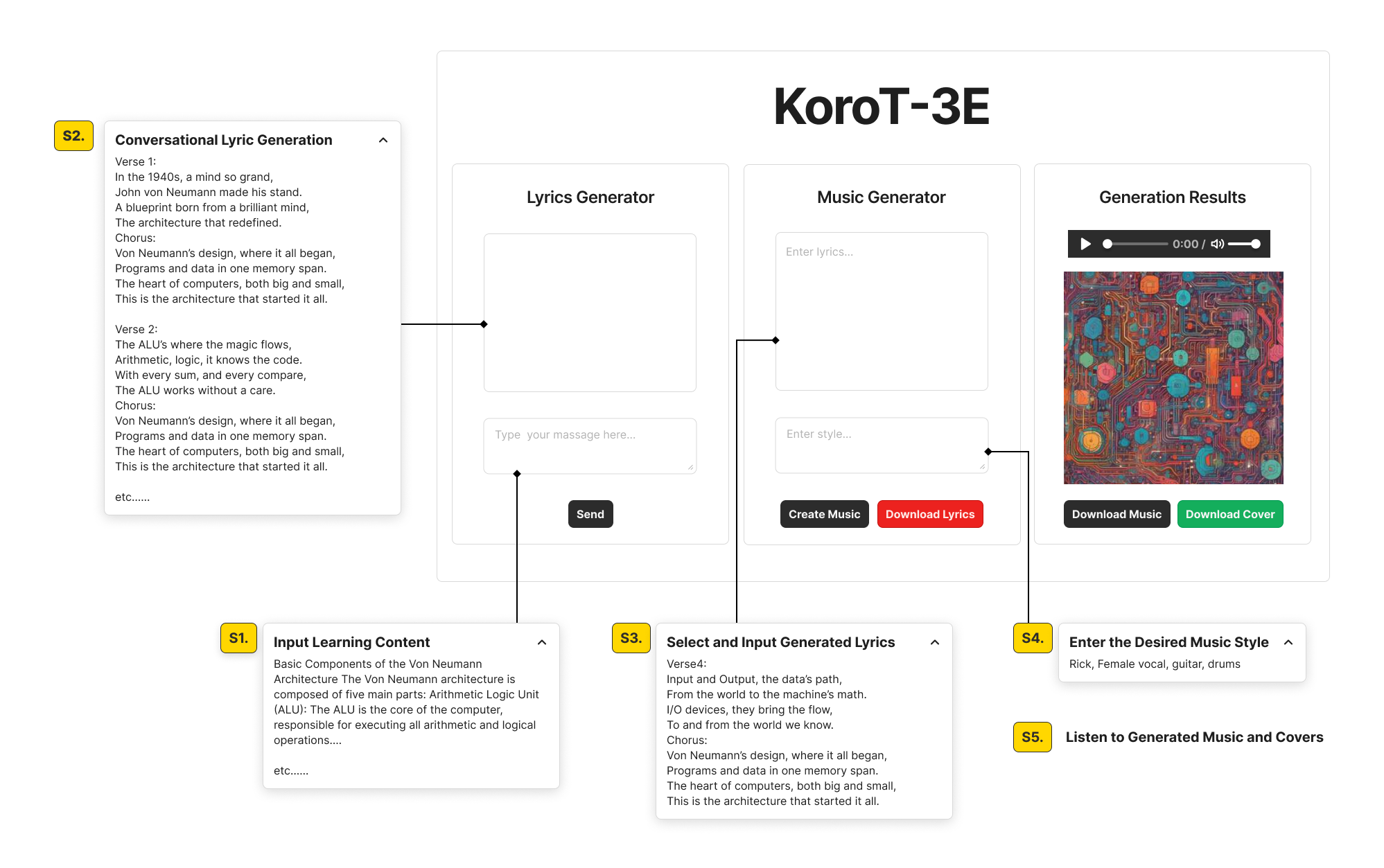}
    \caption{The KoroT-3E interface and usage process.S1: Use the lyrics generator to input the CS concepts that be adapted along with the corresponding prompts. S2 \& S3: Receive the adapted lyrics and copy and paste them into the music generator; the lyrics can also be downloaded as a txt file. S4: Input the user’s preferred music style. S5: Obtain the generated musical mnemonic and music cover, both of which can be downloaded.}
    \label{korot}
\end{figure}

\subsection{Step 1. Lyrics Generation}

We utilize the GPT-4o\footnote{https://openai.com/index/chatgpt/} API for the lyrics generation feature, allowing users to adapt CS concepts into easily memorable and understandable lyrics. As the most powerful and intelligent large language model currently available, GPT-4o can effectively transform concepts into lyrics. Our tests have shown that GPT-4o can adeptly use metaphors and examples to convert abstract and tedious concepts into more engaging lyrics. Additionally, it can flexibly generate lyrics in different styles based on the user's prompt. Users can directly input the name of a concept as a prompt to ask GPT-4o to adapt it into lyrics, or they can input the entire concept content for direct adaptation. However, the first method relies on GPT-4o's own understanding of the concepts and may result in inaccuracies. Therefore, in our experiments, we recommend users adopt the second method.

\subsection{Step 2. Music Generation}

For music generation, we use Suno\footnote{https://suno.com/}, a powerful AI music generator developed by a team from MIT. Users can create high-quality music in various styles by simply providing text prompts without needing any musical background. Suno's ease of use allows users to generate their desired music style with simple prompts. Additionally, Suno can compose music based on user-inputted lyrics, which aligns perfectly with our goal of adapting concepts into lyrics and setting them to music.

We have divided the music generation feature into two modules. The first is the music generation module, where users can input the lyrics adapted by GPT-4o and save the lyrics as a txt document for later reference. Users can also specify their preferred music style here. Once all inputs are completed, users can click the "Create Music" button. The music generated, and a cover image created by Suno based on the lyrics and music style will be available in the Music Generation Results module. Here, users can enjoy and memorize their custom-made musical mnemonics and save the generated music and images for repeated listening and study.

\subsection{Step 3. Music Listening and Saving}

Once the user completes the setup of lyrics and music style, the system automatically proceeds to the music and cover image generation phase. During this stage, the Suno tool generates personalized music based on the provided lyrics and selected music style. Simultaneously, the system creates a cover image inspired by the lyrics, designed by Suno's AI model, enhancing the user's visual experience and adding a personalized touch. Users can directly play the customized music and view the matching cover image within the results module. The generated music and images are available for immediate enjoyment and can be saved to the user's device for future review and repeated listening, aiding in memory retention.

This approach allows users to enjoy a complete and immersive learning experience by transforming concepts into personalized musical mnemonic tools. The design focuses on simplifying operational steps and enhancing usability, ensuring a pleasant and efficient experience for users during repeated use. Additionally, by saving and sharing the generated music and images, users can reinforce their learning outcomes, further improving knowledge retention and mastery.

\section{SYSTEM EVALUATION}

In this section, we present a comprehensive evaluation of the KoroT-3E system to validate its effectiveness and usability in CS educational settings. Through carefully designed experiments, we examine the system's performance in helping novice learners in computer science more effectively memorize and understand fundamental concepts. The evaluation employs a mixed-methods approach, incorporating both quantitative and qualitative data collection and analysis, to provide in-depth insights into how KoroT-3E influences learners' short-term and long-term memory. Our study aims to measure the educational impact of KoroT-3E using empirical data and to identify its potential value and areas for improvement in real-world applications.

\subsection{Participants}
We recruited a group of introductory-level computer science learners to participate in the experiment. Initially, we distributed a pre-survey to assess their knowledge of computer-related topics. From the respondents, we selected 36 participants (n=36) aged between 20 and 30. Most of them frequently used computers and had prior experience with large language models. Their understanding of CS knowledge was relatively uniform. The participants were then evenly divided into an experimental group (n=18) and a control group (n=18). In the experimental group, 7 participants self-identified as female, and 11 as male. In the control group, 8 participants self-identified as female and 10 as male. All participants completed the experiment, and 10 experimental group members agreed to participate in post-experiment interviews. All participants received monetary compensation, with those participating in the interviews receiving additional compensation. The study protocol received approval from our institution's Research Ethics Committee.

\subsection{Experiment Design}

To validate that the finalized KoroT-3E design effectively aids learners in better memorizing and understanding fundamental computer science concepts, we designed an experiment with the assistance of computer education experts. We began by categorizing concepts according to the knowledge dimensions outlined in Anderson and Krathwohl's 2001 revision of Bloom's Taxonomy\cite{anderson2001taxonomy}. We determined that KoroT-3E should be evaluated based on its ability to assist with retaining factual, conceptual, and procedural knowledge. Factual knowledge primarily includes terminology and specific details or elements. This type of knowledge forms the foundation for understanding more complex concepts and principles, often requiring memorization and recall\cite{wilson2016anderson}. Conceptual knowledge encompasses more abstract content, such as classifications, principles, models, and theories, emphasizing the understanding of the structure of knowledge and the relationships between its elements\cite{wilson2016anderson}. Procedural knowledge involves specific steps, techniques, and methods, requiring learners to know what to do and how to do it and to apply these skills in practice\cite{wilson2016anderson}. These three types of knowledge provide a comprehensive assessment of learners' memory and understanding and align well with the difficulties learners face in memorizing and understanding computer science concepts, as identified in our formative study.

After determining the types of concepts to include, we selected the Von Neumann Architecture(VNA)\cite{von1945neumann} and Linear Regression(LR)\cite{su2012linear} as the topics for the experiment. These concepts belong to the domains of CS and data analysis/artificial intelligence, respectively, and are essential for novice learners in these fields with moderate difficulty. The VNA encompasses factual and conceptual knowledge, while LR includes factual, conceptual, and procedural knowledge. This selection allows us to effectively assess how KoroT-3E supports learners in memory and understanding focused knowledge areas. We developed appropriately challenging questions to test the participants. Additionally, to evaluate KoroT-3E's effectiveness in both short-term and long-term memory retention, we referred to the former research\cite{bahrami2019impact,perrault2015physical,yeoh2013musical} and designed the experiment to include tests for both types of memory. Details are below: 

\begin{figure}[t]
    \centering
    \includegraphics[width=0.9\linewidth]{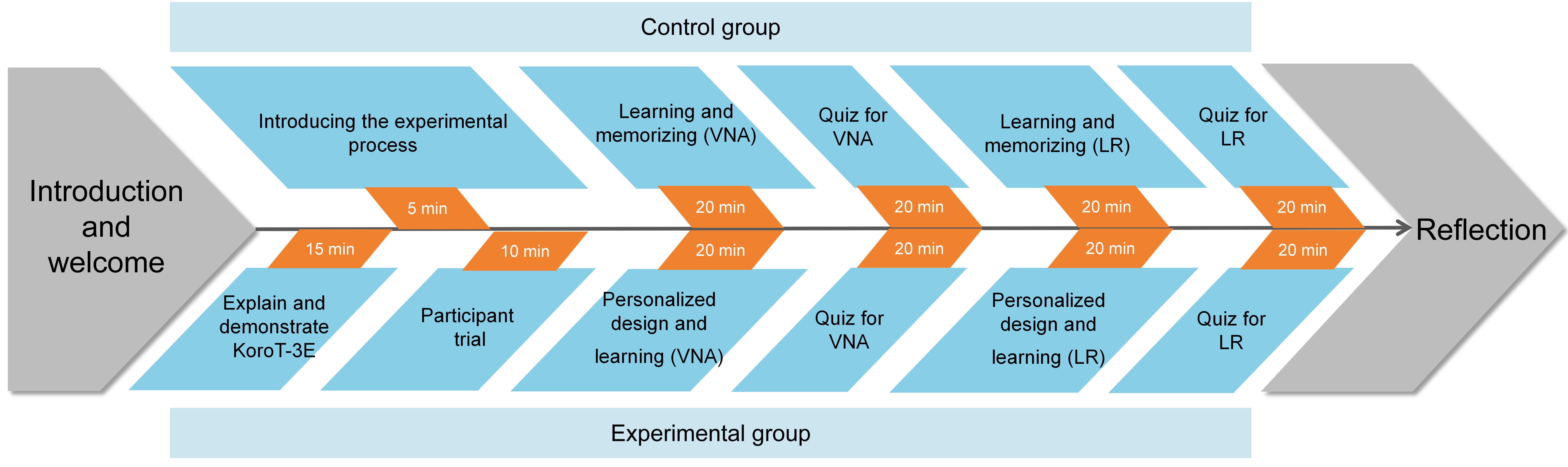}
    \caption{Short-term memory experiment flowchart.}
    \label{experiment1}
\end{figure}

\subsubsection{\textbf{Experiment: Short-term Memory}}
We conducted a controlled experiment, dividing participants into two groups: a control group(n=18) and an experimental group(n=18). The control group was instructed to read and memorize the relevant concepts independently, while the experimental group used KoroT-3E to assist their memorization. The experimental procedure can be referred to \autoref{experiment1}.

Initially, we spent fifteen minutes explaining to the experimental group how to use KoroT-3E, including prompt techniques and other usage tips. They were then given ten minutes to independently explore and familiarize themselves with the system. Once they were comfortable using KoroT-3E, we began the experiment. We prepared a Word document containing the VNA concepts and distributed it to the participants. Each participant was allocated 20 minutes to study and memorize the concepts. During this time, the experimental group used KoroT-3E to adapt these concepts into personalized lyrics and compose suitable melodies based on their musical preferences. After creating their personalized musical mnemonics, they began using them to assist in memorizing and understanding the concepts.
Meanwhile, the control group relied solely on reading the text to memorize and understand the concepts. Throughout the process, if participants had any questions about KoroT-3E or the experimental procedures, we provided immediate assistance. Once the allocated time ended, participants were instructed to stop memorizing, close the concept documents, and were then given a test to assess their understanding of the concepts. After completing the first test, participants were given the concepts of LR and the same process of memorization and testing was repeated within the same time constraints.

\subsubsection{\textbf{Experiment: Long-term Memory}}

To explore the impact of KoroT-3E on long-term memory, we conducted follow-up tests with the participants one day and one week after the initial experiment. These tests aimed to determine whether musical mnemonics facilitated better long-term retention of the concepts. By comparing the test results from these two follow-up periods, we could assess the effectiveness of KoroT-3E in enhancing long-term memory retention compared to traditional study methods.

\subsection{Data collection and analysis}

The test was conducted using the SoJump\footnote{https://www.wjx.cn/}, which incorporates anti-cheating mechanisms and has a fixed exam duration. Two CS experts jointly developed the answer keys and scoring criteria, and all grading was standardized after completing all tests. All questionnaires were collected using Divosurvey\footnote{https://www.divosurvey.com/}, and data analysis was performed using SPSS. The interviews were conducted online, allowing participants to choose between text or verbal responses based on preference. The processing of interview data followed the same approach described in Section 3.1. Throughout the entire process, we made every effort to protect participants' privacy and avoid the excessive collection of personal information.

\section{QUANTITATIVE RESULTS}

\subsection{Participants' performance}

We analyzed each group's performance across the three tests, focusing on the scores for each concept. We report each test's mean, standard deviation, and 95\% confidence interval. An independent samples t-test with an alpha level of 0.05 was used to determine if there was statistical evidence of a significant difference between the mean values of the two groups. Additionally, we analyzed Cohen's effect size to measure the standardized magnitude of the mean differences between the groups. The test results can be referred to \autoref{test}

\begin{figure}[t]
    \centering
    \includegraphics[width=\linewidth]{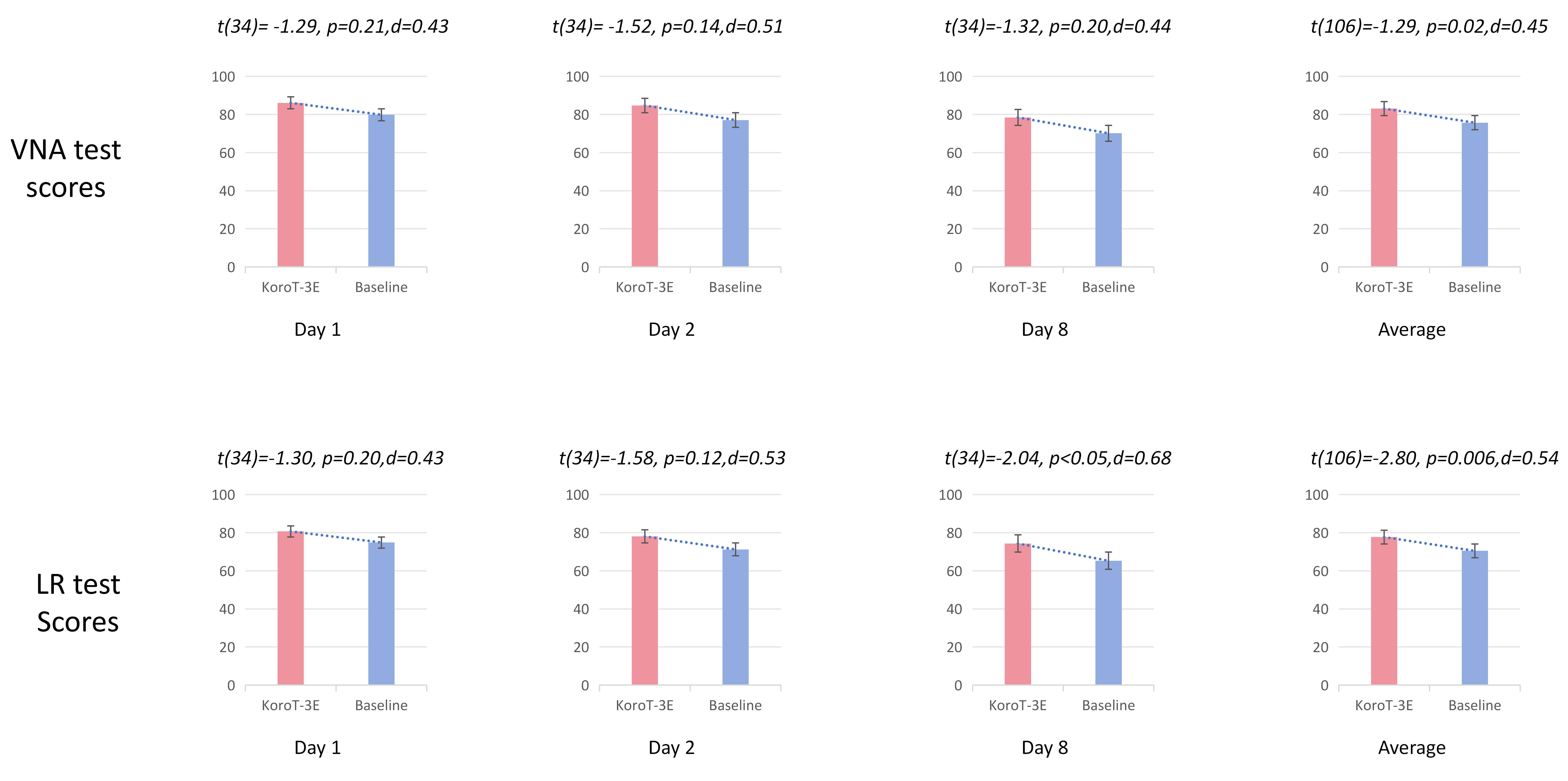}
    \caption{Single test scores and overall average scores for VNA and LR across three tests.}
    \label{test}
\end{figure}

\subsubsection{The test results for VNA}
For VNA, the results of the first test were (Experimental group: M=86.11\%, SD=12.04; Control group: M=79.86\%, SD=16.68; t(34)=-1.29, p=0.21, d=0.43). Although the experimental group had a higher mean score than the control group, the p-value was greater than 0.05, indicating that the difference did not reach statistical significance. The results of the second test for VNA were (Experimental group: M=84.72\%, SD=11.79; Control group: M=77.08\%, SD=17.81; t(34)=-1.52, p=0.14,d=0.51). The difference was not statistically significant. The results of the third test for VNA were (Experimental group: M=78.47\%, SD=18.59; Control group: M=70.14\%, SD=19.24; t(34)=-1.32, p=0.20, d=0.44). The difference remained statistically insignificant.
The overall score for VNA was (Experimental group: M=83.10\%, SD=11.56; Control group: M=75.69\%, SD=18.08; t(106)=-2.34, p=0.02, d=0.45). The p-value was less than 0.05, indicating a significant difference in overall scores between the experimental and control groups. This demonstrates that KoroT-3E had a positive effect on the overall performance of the experimental group in VNA.

\subsubsection{The test results for LR}
For LR, the results of the first test were (Experimental group: M=80.69\%, SD=12.39; Control group: M=74.86\%, SD=14.41; t(34)=-1.30, p=0.20, d=0.43). The p-value was greater than 0.05, indicating that the first test was not statistically significant.
The results of the second test for LR were (Experimental group: M=78.06\%, SD=12.50; Control group: M=71.25\%, SD=13.40; t(34)=-1.58, p=0.12, d=0.53). Again, they were not statistically significant.
The results of the third test for LR were (Experimental group: M=74.31\%, SD=11.94; Control group: M=65.28\%, SD=14.55; t(34)=-2.04, p<0.05, d=0.68). The p-value was slightly less than 0.05, indicating that scores between the experimental and control groups in the third test were statistically significant. This suggests that KoroT-3E favorably impacted the experimental group's performance.
The overall score for LR was (Experimental group: M=77.69\%, SD=12.33; Control group: M=70.46\%, SD=14.42; t(106)=-2.80, p=0.006, d=0.54). The p-value was significantly below 0.05, indicating a significant difference in overall scores between the experimental and control groups. This demonstrates that KoroT-3E also had a notably positive effect on the overall performance of the experimental group in LR.
In the long-term memory test conducted one week later, the control group demonstrated a significant decline in performance. Upon analyzing their responses, we identified that some participants had forgotten certain concepts related to linear regression. More critically, many had forgotten the formulas and the procedural steps required for linear regression calculations, leading to a substantial loss of points.

This observation aligns with cognitive load theory, which suggests that the retention of procedural knowledge requires consistent practice and reinforcement. The decline in performance could be attributed to the lack of opportunities for rehearsal and the absence of contextual cues that facilitate retrieval from long-term memory. Without frequent retrieval practice, learners may struggle with applying complex concepts, particularly when these concepts involve algorithmic thinking or sequential problem-solving steps. Thus, the observed forgetting of formulas and calculation methods significantly impacted the control group’s ability to perform well on the test. The experimental group may have developed stronger long-term memory with the assistance of KoroT-3E. This aspect will be further explored in our qualitative analysis of the interview data.

In summary, KoroT-3E demonstrated significant effectiveness in supporting long-term memory across two concepts and three types of knowledge. Although only one test in LR shows statistically significant results, the experimental group consistently outperformed the control group in terms of average scores. Overall, the experimental group’s performance was significantly better than the control group's throughout the experiment. These findings indicate that KoroT-3E has a positive impact on helping learners memorize and understand concepts.

\subsection{Participants' Cognitive Load and Experimental Experience}

First, we used a 5-point Likert scale to evaluate the cognitive load of participants in both groups, assessing their perceptions of the difficulty in memorizing and understanding the selected concepts within the time constraints of the experiment, as well as their opinions on the design of the test questions. The results showed no statistically significant differences between the two groups (Experimental group: M=2.04/5, Control group: M=2.11/5; p=0.70). No participants reported difficulty in memorizing the concepts within the allotted time (Experimental group: M=1.56/5, Control group: M=1.67/5; p=0.51), nor did they report difficulty in understanding the concepts within the same timeframe (Experimental group: M=1.56/5, Control group: M=1.67/5; p=0.75). Additionally, participants perceived the test difficulty as moderate (Experimental group: M=3.17/5, Control group: M=3.1/5; p=0.87), indicating no objective or subjective differences in cognitive load. This suggests that the selection of concepts, the time allotted for the experiment, and the design of the test questions were all within the participants' learning and comprehension capabilities.

\begin{figure}[htbp]
    \centering
    \includegraphics[width=0.9\linewidth]{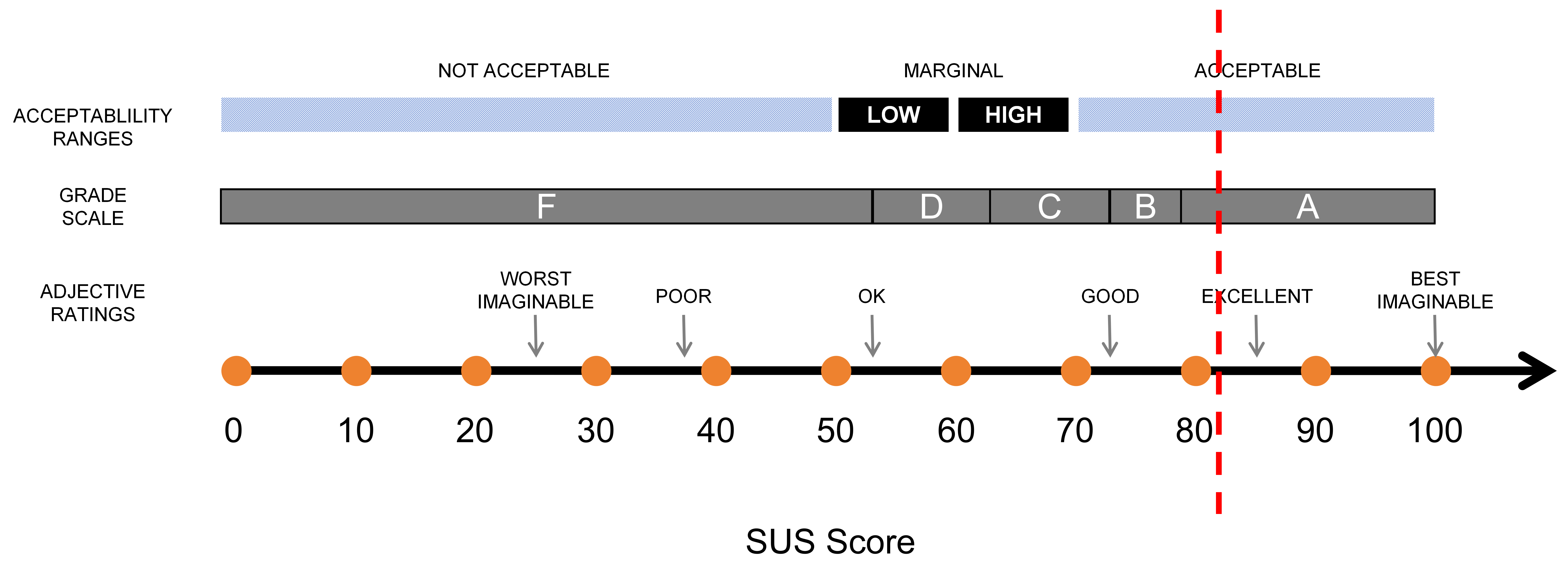}
    \caption{Result of SUS}
    \label{sus}
\end{figure}

For the experimental group, we administered the System Usability Scale (SUS) \cite{jordan1996usability,lewis2018system} after the experiment. The SUS questionnaire consists of 10 items, each rated on a 5-point Likert scale. According to recommendations for interpreting SUS results, a score below 51 is considered failing, while a score above 78.9 corresponds to an "A-" rating. KoroT-3E achieved a final score of 80.4(\autoref{sus}), indicating a highly satisfactory outcome.

Additionally, we employed another 5-point Likert scale (1 being Strongly Disagree, 5 being Strongly Agree) to comprehensively assess the experimental group's perceptions and attitudes towards the practicality of KoroT-3E and whether it aligns with our design theories. The questionnaire covered 11 key aspects, each with several related sub-questions. We calculated the mean and standard deviation for each aspect: perceived usefulness (M=3.7/5, SD=0.83), perceived ease of use (M=4.2/5, SD=0.51), usage attitude (M=4.1/5, SD=0.42), usage intention(M=3.6/5, SD=1.09), interest in KoroT-3E (M=4.3/5, SD=0.60), Continuous usage (M=3.7/5, SD=1.04), attractiveness (M=3.9/5, SD=0.87), novelty (M=4.3/5, SD=0.39), constructivist learning theory experience (M=4.1/5, SD=0.75), and experiential learning theory experience (M=3.9/5, SD=0.84). For detailed information about the scale, please refer to \autoref{scale} in the appendix. We also observed the musical styles chosen by participants, noting that each participant set different prompts. The preferred musical styles varied for each individual.

\section{QUALITATIVE RESULTS}
After completing the long-term memory experiment, we invited members of the experimental group to participate in interviews to gain deeper insights into their perspectives and suggestions. Ultimately, 10 participants (n=10)agreed to participate in the interviews. Our interviews were conducted in a semi-structured format to gain an in-depth understanding of the perspectives and attitudes of participants, particularly novice learners, toward using musical mnemonics to aid memory retention. Our analysis offers valuable insights into the participants' subjective experiences, their acceptance of the KoroT-3E design concept, and their recognition of this musical mnemonic approach as a tool for enhancing memory. Detailed information about the interview participants can be found in \autoref{tab:participants}.

\begin{table}[h]
\centering
\caption{Interview Participants' Information}
\begin{tabular}{>{\centering\arraybackslash}m{1.5cm}|>{\centering\arraybackslash}m{2cm}|>{\centering\arraybackslash}m{2cm}|>{\centering\arraybackslash}m{4cm}}
\hline
\textbf{ID} & \textbf{Age} & \textbf{Gender} & \textbf{Field of Study/Work} \\ \hline
p1 & 20-30 & F & Marketing \\ \hline
p2 & 20-30 & F & Geological \\ \hline
p3 & 20-30 & F & Internet \\ \hline
p4 & 20-30 & M & Bank \\ \hline
p5 & 20-30 & F & Medical \\ \hline
p6 & 20-30 & M & Operation \\ \hline
p7 & 20-30 & M & Business \\ \hline
p8 & 20-30 & M & Engineer \\ \hline
p9 & 20-30 & M & Financial \\ \hline
p10 & 20-30 & F & Electrical \\ \hline
\end{tabular}
\label{tab:participants}
\end{table}

\subsection{Feature and Usability Feedback}

\subsubsection{The interface design is simple and easy to use.}

While we utilized the SUS scale to assess user evaluations of KoroT-3E's usability, we also sought to gain a deeper understanding of their attitudes and opinions regarding the overall functionality and interface design.
For example, P1 mentioned, \textit{``I find the interface quite user-friendly, with all the functions clearly visible. It should be easy to use even for those not proficient with computers." (P1)}. Similarly, P7 noted that the interface is very easy to navigate: \textit{``"First, I think the interface is relatively simple, allowing me to quickly find the necessary functions. It's visually appealing, with clearly defined sections, and I didn't encounter any difficulties while using it. I could easily operate it."(P7)}.

\subsubsection{The lyrics generation functionality delivers impressive results.}
One of the key features of KoroT-3E is its ability to automatically adapt concepts into lyrics, a functionality that received positive feedback from many participants. As P2 stated:
\begin{quote}
\textit{``It not only creates the final musical composition but also provides lyrics that can be followed while listening to the music. This way, users can memorize the lyrics and then use the music to remember the concepts. Including lyrics lowers the barrier to listening to the music, enhances the software's accessibility, and makes it usable for everyone. "(P2)}
\end{quote}

Additionally, writing good lyrics is often challenging for most people since it requires attention to rhythm and rhyme. Transforming concepts into lyrics adds another layer of difficulty.P3 also praised this feature: 
\textit{``In Chinese, generating lyrics requires considering their actual meaning, rhythm, and rhyme, which is honestly quite challenging. Considering the function of the lyrics, if they are well-constructed, the user might not even need the accompanying music—they could mentally sing the lyrics themselves. So, I think the lyric generation feature is quite good."(P3)}.

\subsubsection{Enables users to customize the musical style.}
To align with constructivist learning theory\cite{hein1991constructivist} and generation effect\cite{slamecka1978generation} and provide a more personalized learning experience, we offer users the ability to generate their own music, creating music mnemonics that match their preferred musical styles. 
As P5 and P9 noted:
\textit{``The ability to adapt to different musical styles is great because everyone has different preferences. If the song's style isn't to their liking, they might not be interested in using it."(P5)} and 
\textit{``Creating your own song allows the style to better match your preferences, making it more enjoyable to listen to. It also lets you focus on specific areas, making the song more tailored to your needs."(P9)}.

\subsection{Learning Effectiveness and Memory Retention}

\subsubsection{Improves the efficiency of memorization.}

Music education can enhance students' cognitive abilities, academic performance, and memory\cite{yang2023impact,hodges2005impact,scripp2002overview}. Musical activities, such as playing an instrument or singing, help develop transferable skills that benefit other subjects like mathematics, science, and languages \cite{scripp2002overview}. This context is known as Transfer of Learning\cite{perkins1992transfer}. According to participants, using musical mnemonics to memorize concepts also makes the memorization process more efficient. For example, P4 mentioned, 
\textit{``With lyrics and melody, the concepts become easier to remember. There are a couple of classic melodies that I can still sing even now. Melodies from songs are indeed easier and quicker to memorize than plain concepts." (P4)}. P7 shared a similar view: \textit{``The repetition of some keywords matched with the rhythm and melody of the music helps me repeatedly recall a particular concept, making it easier to remember."  (P7)}. Music education is also believed to facilitate the formation of long-term memory \cite{groussard2010music}. P2 agreed, stating, 
\begin{quote}
\textit{``In terms of efficiency, it's much easier than rote memorization of a concept. I think its greater value lies in facilitating long-term memory; through repeated listening, concepts are more likely to be retained in the long term."(P2)}
\end{quote}

\subsubsection{Enhances memory retention.}

Research has shown that music can aid in memory retention\cite{roden2012effects,alexomanolaki2007music}. Musical mnemonics build upon this by integrating concepts into lyrics, offering learners a more engaging learning experience. As P6 shared, \textit{``The arrangement of the lyrics has a certain rhythm, making them catchy and easy to recite or sing. This makes the process feel less tedious compared to traditional rote memorization and knowledge recall."  (P6)}. An earworm refers to a fragment of music that involuntarily repeats in one's mind.\cite{halpern2011persistence} As a common phenomenon, it has been widely used in the commercial sector to strengthen brand associations and enhance the overall impact of advertising by creating consistent memory links\cite{das2023musical}. Our participants reported experiencing a similar effect with KoroT-3E. P5 stated, \begin{quote}
\textit{``The musical elements anchor the concepts. If you listen to a song repeatedly, you'll remember some of its catchiest parts. For instance, this time, I got 'hooked' on a line or two, which helped me recall-related points nearby. The most memorable parts were often the simple, repetitive sequences. "(P5)}
\end{quote}
This earworm effect provides learners with longer-lasting associative memory, where they might even find themselves humming certain segments unintentionally. P9 and P10 echoed these sentiments: \textit{``Songs differ greatly from ordinary recordings. For example, in our verses and choruses, some keywords are repeated, and this repetition is combined with the rhythm of the music. It’s like some viral songs that get stuck in your head—you might find yourself unconsciously humming them in daily life." (P9)} and 
\textit{``Even after the experiment, I still remember a few tunes, which sometimes play back in my mind. I can even vividly recall the AI-generated male voice." (P10)}.

\subsubsection{Facilitates the Understanding of concepts.}
Breaking down concepts to make them easier to understand and remember is known as "chunking." By dividing large amounts of information into smaller, more manageable "chunks," this method reduces cognitive load and enhances memory retention\cite{thalmann2019does}. It helps organize complex information into meaningful units, making it easier to comprehend and recall. For participants, the way KoroT-3E transforms knowledge into lyrics effectively employs chunking. As P2 stated, 
\begin{quote}
\textit{``My original memorization was linear, rote learning—memorizing word by word from start to finish. After the knowledge was turned into lyrics and set to music, it was restructured and presented in a simpler and more understandable way without losing the original content." (P2)}
\end{quote}

P3 expressed a similar view, noting that the melody of the music further reinforced the memory and understanding of the concepts chunked into lyrics. 
\textit{``The adapted lyrics deconstruct and reorganize the concepts, while the melody enhances the retention of these lyrics—or concepts—in an enjoyable and engaging format, making the learning experience simpler and more accessible." (P3)}. In information processing theory, memory and understanding are closely interconnected. Effective comprehension requires storing information in long-term memory, and the depth and retention of memory also depend on the learner's level of understanding\cite{simon1978information}. As P6 noted,
\textit{``Transforming concepts into music helps me better remember and understand complex concepts by simplifying and making them more conversational, making the content easier to read and understand. The addition of rhythm encourages me to sing along, enhancing comprehension and memory retention." (P6)}. 
P9 also observed that 
\textit{``Music aids in memorizing these concepts, which, in turn, positively reinforces their understanding." (P9)}

\subsection{Emotional and Interest Response}

\subsubsection{Engaging and Novel, Increases Learning Interest}

Research has shown that interest is a strong motivator for student learning\cite{ainley2002interest,renninger2014role}. In our formative study, we also found that one of the challenges learners without a background in computer science face when starting to learn foundational computer concepts is a lack of interest. However, after using KoroT-3E, many participants reported that this novel learning approach gave them a surprising experience and increased their interest in learning. P1 stated that \textit{``It brought a novel experience and offered me a new point of interest in learning."(P1)}Furthermore, integrating music as a multimodal learning approach added an element of enjoyment to the otherwise monotonous memorization process. P8 observed that

\begin{quote}
\textit{``Listening to songs repeatedly helps reinforce the memory of concepts without quickly becoming tedious. In contrast, repeatedly reviewing traditional concepts can often feel dull and boring."(P8)}
\end{quote}

\subsubsection{Enhances Learner Initiative and Immersion in Learning.}

A key aspect of constructivist learning theory is the emphasis on learner agency\cite{hein1991constructivist}, which is also one of our design principles. Participants have reported that KoroT-3E indeed fosters greater learning initiative. P5 mentioned, 
``This new method undoubtedly sparked my curiosity, making me more willing to spend time on otherwise uninteresting concepts because I wanted to understand and appreciate this entirely new way of presenting knowledge."(P5)}Every learner has their own preferences, and some essential yet tedious information can create pressure, leading to a sense of forced learning\cite{shuang2018forced}. P2 highlighted that KoroT-3E offers a solution to this problem: 
\textit{``We tend to gravitate toward learning content that we find more interesting, but sometimes we must learn some dull yet useful concepts. In such situations, learning can feel somewhat compulsory. However, with this tool, we can transform content we don't naturally find engaging into something we do, like music we enjoy. This approach makes us more proactive in our learning."(P2)} 
Creating an appropriate learning environment is essential for learners\cite{mash2020creating}, and the Multisensory Learning Theory suggests that multisensory training can be more effective than single-sensory training because it aligns more closely with the way the brain naturally learns in real-world environments.\cite{shams2008benefits}. Participants found that, with the help of suitable music, KoroT-3E also helped them immerse themselves more quickly and effectively in the learning atmosphere. P3 stated, 
\begin{quote}
\textit{``The combination of text and music is multimodal; you have to look at the lyrics while also listening to the melody. This dual stimulation of vision and hearing makes the brain more excited and helps people more easily enter that atmosphere, becoming more immersed."(P3)}
\end{quote} 
P8 added that music helps them feel more relaxed, making it easier to enter a learning state: \textit{``Listening to music makes me feel more relaxed, and in a more relaxed state, it’s easier to start learning."(P8)}

\subsubsection{Elicits Positive Emotions.}

Positive emotions can enhance learning motivation and efficiency\cite{mega2014makes,nie2023research}. Most participants felt that KoroT-3E brought them positive emotions. P6 mentioned, 
\textit{``I feel more positive about learning; this method is more relaxed and entertaining, which has increased my interest in the concepts. The most important aspect is that songs naturally help people feel more at ease."(P6)}

P2 offered an interesting perspective, suggesting that KoroT-3E creates a new dynamic for learners: 
\begin{quote}
\textit{``I think most people feel some degree of positive emotion when they hear music they like. Since the lyrics of this music are composed of concepts, it extends people's positive emotions about the music to the knowledge itself, which could lead to a more positive attitude toward learning."(P2)}
\end{quote}

From his viewpoint, the positive emotions derived from listening to his favorite music are conveyed through the lyrics, subtly fostering a more positive emotional response to learning.

\subsection{Reflection on Learning Methods and New Perceptions}

\subsubsection{Surpasses Traditional Memory Methods.}

Although various mnemonic techniques, such as traditional mnemonics, have been popular for many years, participants found that KoroT-3E, which integrates advanced technology to create AI-generated mnemonics, brought them many pleasant surprises. As P10 noted,
\textit{``Back when I was preparing for my graduate exams, I learned a lot of mnemonic songs online, but some weren't enjoyable. With KoroT-3E, I can now generate music that I like."(P10)} 
This innovative approach also led participants to reconsider their future learning strategies. P8 mentioned, 
\begin{quote}
\textit{``I used to rely on rote memorization, which was time-consuming and ineffective. Using a method like this—where concepts are turned into songs and set to a preferred musical style—not only enhances learning efficiency but also offers relaxation."(P8)}
\end{quote} 

\subsubsection{Advocates for Increased Use of Technology in Learning.}

Although the participants were already familiar with using cutting-edge technologies like LLMs, KoroT-3E's application of these technologies to learning gave them new perspectives on how to use technology effectively. 
As P6 expressed, 
\textit{``We should explore new ways to memorize and recall concepts instead of sticking to traditional methods. I believe we need more innovation in future learning approaches."(P6)}

P4 added, 
\begin{quote}
\textit{``I think there should be more inclusivity and exploration of new channels in learning methods. This is one of the main reasons I find KoroT-3E practical; it offers fresh ideas and changes how we might approach learning in the future."(P4)}
\end{quote}

\subsection{Applicability and Potential Users}

To explore KoroT-3E's broader applicability across different disciplines, fields, and user groups, we sought participants' opinions. They all responded positively and provided specific insights and suggestions based on their own areas of study or work.

Participants generally believed that KoroT-3E could be widely applied to disciplines or fields with substantial memorization requirements, such as medicine, law, and social sciences. Regarding potential use cases, P1 mentioned an interesting scenario beyond education: \textit{applying this concept to marketing by turning advertising slogans into earworms, enhancing their reach and impact.}

When discussing target audiences, participants suggested a wide range. However, P7 raised an intriguing point: 

\begin{quote}
\textit{``I don't think it's suitable for children. For children, it could be challenging for them to identify errors in the generated songs, potentially leading to confusion about the concepts. Since we cannot control the AI generation process, children, who lack the basic ability to recognize or understand errors, might develop misconceptions due to such a product." (P4)}
\end{quote}

\subsection{Limitations}

The most frequently mentioned limitation by participants is the unpredictability of adapting lyrics using LLMs. Despite using the most advanced GPT-4o model and employing prompt engineering to ensure the accuracy and quality of the concept adaptation into lyrics, the results still exhibit instability in practice. Issues include overlooking certain concepts, incorporating illogical metaphors or parallel structures, or generating long, complex sentences that are difficult to parse.

Another limitation, as highlighted by P1,
\begin{quote}
\textit{``I don’t listen to music often, so I’m not sure what type of song suits me. The first song I generated felt noisy and distracted me, but things improved when I tried a different style."(P1)}
\end{quote}
This points to a broader concern: many people may not have the habit of listening to music regularly, especially while studying. Consequently, they may not be familiar with different music styles or know which style best suits them, potentially causing confusion or dissatisfaction.

\section{DISCUSSION AND FUTURE WORK}

\subsection{Musical Mnemonics for CS Learning}

As a technique that has existed for thousands of years and is still in use today, the effectiveness of mnemonics is indisputable. Musical mnemonics, as a multimodal memory aid, enhance information processing and retention through multiple sensory channels with the help of melodies and rhythms. This has been partially confirmed in our experiments: overall, both the VNA and LR experimental groups using KoroT-3E for memory support outperformed the control group. Although none of the single-test scores, except for the third LR test, showed statistical significance, the average scores of the experimental group were still higher than those of the control group. In the third LR test, a notable decline in the control group's performance was observed, which aligns with cognitive load theory. As procedural knowledge, if retrieval practice is not conducted regularly, learners may struggle with applying complex concepts, particularly when these concepts involve algorithmic thinking or sequential problem-solving steps. With the assistance of musical mnemonics, the experimental group members could better remember these concepts and steps.

The results for all participants showed that our choice of concepts, experimental design, and test items were well-suited to the memory and comprehension levels of these introductory-level CS participants. Participants found KoroT-3E easy to use, with intuitive functionality and a low learning curve, receiving positive evaluations. Additionally, the varied music styles chosen by each participant highlighted diverse musical preferences, underscoring the need for personalization.

During the interviews, participants further discussed their perspectives on KoroT-3E. They noted that the personalized musical mnemonics generated by KoroT-3E helped them memorize information more efficiently, with the phenomenon of "earworms" enhancing memory retention. Adapting concepts into lyrics by KoroT-3E involved a chunking-like step, which simplified complex information structures and made the learning material easier to understand and remember. Moreover, musical mnemonics, as an innovative multimodal learning method, aligned with Multisensory Learning Theory, increased their interest in learning, promoted active engagement, and created a positive emotional atmosphere and improved learning environment. Overall feedback suggests that KoroT-3E successfully achieved the design principles we established based on constructivist learning theory. It also provided a potential solution to the challenges identified in our formative study related to novice learners' difficulties in retaining fundamental concepts in CS.

Furthermore, participants had a highly positive attitude towards the applicability and potential users of KoroT-3E. They all believed that KoroT-3E could be applied to more subjects and broader fields beyond education. However, given the current issues with AI technology, such as its black-box nature, hallucinations, and various potential biases, our ongoing challenge is to find ways to make AI applications in education more effective, stable, and safe.

\subsection{Advanced Technology for Learning}

With the rise of new technologies in recent years, researchers have explored their integration into education, revitalizing learning experiences. However, many of these methods have not achieved widespread adoption. Gaps between disciplines often lead to educators and learners lacking awareness of the latest technologies and their potential applications\cite{margulieux2024recommendations}.

KoroT-3E provides an example of how advanced technology can be applied to novel learning domains, yielding promising results. To foster more effective educational tools, it is crucial to enhance interdisciplinary and cross-field collaboration. By bringing together experts from various domains, such as cognitive science, computer science, and education. We can ensure that these tools are not only technologically advanced but also pedagogically sound and psychologically effective. For instance, insights from HCI can help design user-friendly interfaces that enhance learner engagement, while contributions from cognitive science can refine the underlying models to align with how the human brain processes and retains information.

In summary, fostering such collaboration can lead to the development of innovative educational tools that are effective and widely accepted. This collaborative approach will enable us to leverage the potential of emerging technologies in education, ultimately enriching the learning experiences and outcomes for all students.

\subsection{Design and Research Implication}

The development and evaluation of KoroT-3E highlight several key factors to consider when designing future educational tools to enhance the learning of CS concepts. Firstly, personalization is an essential factor that cannot be overlooked. Many interview participants indicated that music styles aligned with their preferences were particularly helpful. Future designs for learning support tools should further explore personalization to create more tailored learning experiences, resulting in better learning outcomes.

Secondly, KoroT-3E's success in using musical mnemonics indicates the potential for diversifying mnemonic strategies in educational technology. Future research should explore integrating other mnemonic techniques, such as visual aids, storytelling, or gamification, to accommodate different learning styles and contexts. This approach can not only enrich the learning experience but also expand the applicability of mnemonic tools beyond traditional settings.

Additionally, this study emphasizes the potential for extending the application of multimodal AI-based tools beyond computer science to various other fields. Disciplines that require extensive memorization, such as medicine, law, and social sciences, could benefit from similar tools customized to their specific content and learning challenges. Insights from human-computer interaction, cognitive psychology, and educational theories should be integrated to refine the design and application of such tools.

\subsection{Limitations}
\subsubsection{Limitations of AI in Education}
Despite the rapid advancement of generative AI technology, significant limitations remain. For example, large language models (LLMs) can hallucinate and generate non-existent content \cite{li2024dawn}, and biases or errors in training data can still affect users. While prompt engineering can mitigate some of these issues, it cannot fully eliminate them. Additionally, although Suno is one of the most advanced music generation models, it struggles with certain vocal issues, such as incorrect or unclear pronunciation. While adults may overcome these problems by generating multiple outputs until satisfactory, for less discerning populations, like children, this could hinder learning and have serious consequences \cite{lewandowsky2012misinformation}. This concern is why Participant P4 finds the software unsuitable for children. As AI expands in education, establishing effective interdisciplinary partnerships among AI developers, educators, and researchers is crucial\cite{giannakos2024promise}. Such collaboration ensures AI technologies meet educational needs, help educators understand AI's potential and limitations, and safeguard ethical standards in AI applications.

\subsubsection{Limitations of Interaction}

During expert consultation, some experts suggested adding a scrolling lyrics feature to enhance user experience, similar to music players. However, since Suno generates music directly from lyrics without a timeline, attempts to create one using speech recognition models have been unsuccessful. The most accurate method, manually setting the timeline, is impractical for KoroT-3E. Therefore, we cannot currently implement scrolling lyrics but will continue exploring solutions. Additionally, while Participant P1 noted that selecting a suitable style might be challenging for those less familiar with music, KoroT-3E supports repeated and varied music generation, allowing users to experiment until they find a preferred style, minimizing the issue over time.

\section{CONCLUSION}

This study leverages cutting-edge LLMs and music generation models to create personalized musical mnemonics, offering a more effective memory-enhancing learning tool for novice CS learners. The results show that the experimental group performed significantly better than the control group. Surveys and interviews indicated that participants found KoroT-3E easy to use and believed it improved memory retention, increased memory efficiency, and boosted their interest and motivation to learn. Participants also felt that KoroT-3E could be applied to a broader range of fields and user groups. Additionally, we analyzed our concept from the perspective of HCI practitioners, providing new insights into applying advanced technologies such as generative AI in the educational domain. We hope this research inspires further exploration of tools that support foundational computer science knowledge and encourages greater integration of technology across various educational fields.

\bibliographystyle{ACM-Reference-Format}
\bibliography{refs}


\begin{thebibliography}{93}


\ifx \showCODEN    \undefined \def \showCODEN     #1{\unskip}     \fi
\ifx \showDOI      \undefined \def \showDOI       #1{#1}\fi
\ifx \showISBNx    \undefined \def \showISBNx     #1{\unskip}     \fi
\ifx \showISBNxiii \undefined \def \showISBNxiii  #1{\unskip}     \fi
\ifx \showISSN     \undefined \def \showISSN      #1{\unskip}     \fi
\ifx \showLCCN     \undefined \def \showLCCN      #1{\unskip}     \fi
\ifx \shownote     \undefined \def \shownote      #1{#1}          \fi
\ifx \showarticletitle \undefined \def \showarticletitle #1{#1}   \fi
\ifx \showURL      \undefined \def \showURL       {\relax}        \fi
\providecommand\bibfield[2]{#2}
\providecommand\bibinfo[2]{#2}
\providecommand\natexlab[1]{#1}
\providecommand\showeprint[2][]{arXiv:#2}

\bibitem[Agramonte and Belfiore(2002)]%
        {agramonte2002using}
\bibfield{author}{\bibinfo{person}{Valerie Agramonte} {and} \bibinfo{person}{Phillip~J Belfiore}.} \bibinfo{year}{2002}\natexlab{}.
\newblock \showarticletitle{Using mnemonics to increase early literacy skills in urban kindergarten students}.
\newblock \bibinfo{journal}{\emph{Journal of Behavioral Education}} \bibinfo{volume}{11}, \bibinfo{number}{3} (\bibinfo{year}{2002}), \bibinfo{pages}{181--190}.
\newblock


\bibitem[Ainley et~al\mbox{.}(2002)]%
        {ainley2002interest}
\bibfield{author}{\bibinfo{person}{Mary Ainley}, \bibinfo{person}{Suzanne Hidi}, {and} \bibinfo{person}{Dagmar Berndorff}.} \bibinfo{year}{2002}\natexlab{}.
\newblock \showarticletitle{Interest, learning, and the psychological processes that mediate their relationship.}
\newblock \bibinfo{journal}{\emph{Journal of educational psychology}} \bibinfo{volume}{94}, \bibinfo{number}{3} (\bibinfo{year}{2002}), \bibinfo{pages}{545}.
\newblock


\bibitem[Akpan et~al\mbox{.}(2021)]%
        {akpan2021impact}
\bibfield{author}{\bibinfo{person}{Joseph Akpan}, \bibinfo{person}{Charles~E Notar}, {and} \bibinfo{person}{Larry Beard}.} \bibinfo{year}{2021}\natexlab{}.
\newblock \showarticletitle{The impact of mnemonics as instructional tool}.
\newblock \bibinfo{journal}{\emph{Journal of Education and Human Development}} \bibinfo{volume}{10}, \bibinfo{number}{3} (\bibinfo{year}{2021}), \bibinfo{pages}{20--28}.
\newblock


\bibitem[Alexomanolaki et~al\mbox{.}(2007)]%
        {alexomanolaki2007music}
\bibfield{author}{\bibinfo{person}{Margarita Alexomanolaki}, \bibinfo{person}{Catherine Loveday}, {and} \bibinfo{person}{Chris Kennett}.} \bibinfo{year}{2007}\natexlab{}.
\newblock \showarticletitle{Music and memory in advertising: Music as a device of implicit learning and recall}.
\newblock \bibinfo{journal}{\emph{Music, Sound, and the Moving image}} \bibinfo{volume}{1}, \bibinfo{number}{1} (\bibinfo{year}{2007}), \bibinfo{pages}{51--71}.
\newblock


\bibitem[Anderson and Krathwohl(2001)]%
        {anderson2001taxonomy}
\bibfield{author}{\bibinfo{person}{Lorin~W Anderson} {and} \bibinfo{person}{David~R Krathwohl}.} \bibinfo{year}{2001}\natexlab{}.
\newblock \bibinfo{booktitle}{\emph{A taxonomy for learning, teaching, and assessing: A revision of Bloom's taxonomy of educational objectives: complete edition}}.
\newblock \bibinfo{publisher}{Addison Wesley Longman, Inc.}
\newblock


\bibitem[{\"A}ngeslev{\"a} et~al\mbox{.}(2003)]%
        {angesleva2003body}
\bibfield{author}{\bibinfo{person}{Jussi {\"A}ngeslev{\"a}}, \bibinfo{person}{Ian Oakley}, \bibinfo{person}{Stephen Hughes}, {and} \bibinfo{person}{Sile O’Modhrain}.} \bibinfo{year}{2003}\natexlab{}.
\newblock \showarticletitle{Body Mnemonics Portable device interaction design concept}. In \bibinfo{booktitle}{\emph{Proceedings of UIST}}, Vol.~\bibinfo{volume}{3}. Citeseer, \bibinfo{pages}{2--5}.
\newblock


\bibitem[Arawjo et~al\mbox{.}(2017)]%
        {arawjo2017teaching}
\bibfield{author}{\bibinfo{person}{Ian Arawjo}, \bibinfo{person}{Cheng-Yao Wang}, \bibinfo{person}{Andrew~C Myers}, \bibinfo{person}{Erik Andersen}, {and} \bibinfo{person}{Fran{\c{c}}ois Guimbreti{\`e}re}.} \bibinfo{year}{2017}\natexlab{}.
\newblock \showarticletitle{Teaching programming with gamified semantics}. In \bibinfo{booktitle}{\emph{Proceedings of the 2017 CHI conference on human factors in computing systems}}. \bibinfo{pages}{4911--4923}.
\newblock


\bibitem[Azmi et~al\mbox{.}(2016)]%
        {azmi2016case}
\bibfield{author}{\bibinfo{person}{Mohd Nazri~Latiff Azmi}, \bibinfo{person}{Muhammad Hadi Syafiq~Mohd Najmi}, {and} \bibinfo{person}{Nurazan~Mohmad Rouyan}.} \bibinfo{year}{2016}\natexlab{}.
\newblock \showarticletitle{A case study on the effects of mnemonics on English vocabulary}.
\newblock \bibinfo{journal}{\emph{International Journal of Applied Linguistics and English Literature}} \bibinfo{volume}{5}, \bibinfo{number}{7} (\bibinfo{year}{2016}), \bibinfo{pages}{178--185}.
\newblock


\bibitem[Bada and Olusegun(2015)]%
        {bada2015constructivism}
\bibfield{author}{\bibinfo{person}{Steve~Olusegun Bada} {and} \bibinfo{person}{Steve Olusegun}.} \bibinfo{year}{2015}\natexlab{}.
\newblock \showarticletitle{Constructivism learning theory: A paradigm for teaching and learning}.
\newblock \bibinfo{journal}{\emph{Journal of Research \& Method in Education}} \bibinfo{volume}{5}, \bibinfo{number}{6} (\bibinfo{year}{2015}), \bibinfo{pages}{66--70}.
\newblock


\bibitem[Bahrami et~al\mbox{.}(2019)]%
        {bahrami2019impact}
\bibfield{author}{\bibinfo{person}{Zahra~Nikkhah Bahrami}, \bibinfo{person}{Siros Izadpanah}, {and} \bibinfo{person}{Houman Bijani}.} \bibinfo{year}{2019}\natexlab{}.
\newblock \showarticletitle{The Impact of Musical Mnemonic on Vocabulary Recalling of Iranian Young Learners.}
\newblock \bibinfo{journal}{\emph{International Journal of Instruction}} \bibinfo{volume}{12}, \bibinfo{number}{1} (\bibinfo{year}{2019}), \bibinfo{pages}{977--994}.
\newblock


\bibitem[Balaji et~al\mbox{.}(2013)]%
        {balaji2013mnemonics}
\bibfield{author}{\bibinfo{person}{Dr~RD Balaji}, \bibinfo{person}{Dr~Brijesh Ramniklal}, \bibinfo{person}{N Balasupramanian}, {and} \bibinfo{person}{Er~R Malathi}.} \bibinfo{year}{2013}\natexlab{}.
\newblock \showarticletitle{Mnemonics for higher education using contemporary technologies}.
\newblock \bibinfo{journal}{\emph{arXiv preprint arXiv:1305.2609}} (\bibinfo{year}{2013}).
\newblock


\bibitem[Below({[n.\,d.]})]%
        {belowguide}
\bibfield{author}{\bibinfo{person}{Activity Based Approach Are~Listed Below}.} \bibinfo{year}{[n.\,d.]}\natexlab{}.
\newblock \showarticletitle{Guide To Teaching Computer Science An Activity Based Approach}.
\newblock  (\bibinfo{year}{[n.\,d.]}).
\newblock


\bibitem[Bertsch et~al\mbox{.}(2007)]%
        {bertsch2007generation}
\bibfield{author}{\bibinfo{person}{Sharon Bertsch}, \bibinfo{person}{Bryan~J Pesta}, \bibinfo{person}{Richard Wiscott}, {and} \bibinfo{person}{Michael~A McDaniel}.} \bibinfo{year}{2007}\natexlab{}.
\newblock \showarticletitle{The generation effect: A meta-analytic review}.
\newblock \bibinfo{journal}{\emph{Memory \& cognition}}  \bibinfo{volume}{35} (\bibinfo{year}{2007}), \bibinfo{pages}{201--210}.
\newblock


\bibitem[Bower(1970)]%
        {bower1970analysis}
\bibfield{author}{\bibinfo{person}{Gordon~H Bower}.} \bibinfo{year}{1970}\natexlab{}.
\newblock \showarticletitle{Analysis of a mnemonic device: Modern psychology uncovers the powerful components of an ancient system for improving memory}.
\newblock \bibinfo{journal}{\emph{American Scientist}} \bibinfo{volume}{58}, \bibinfo{number}{5} (\bibinfo{year}{1970}), \bibinfo{pages}{496--510}.
\newblock


\bibitem[Campabello et~al\mbox{.}(2002)]%
        {campabello2002music}
\bibfield{author}{\bibinfo{person}{Nicolette Campabello}, \bibinfo{person}{Mary~Jane De~Carlo}, \bibinfo{person}{Jean O'neil}, {and} \bibinfo{person}{Mary~Jill Vacek}.} \bibinfo{year}{2002}\natexlab{}.
\newblock \showarticletitle{Music Enhances Learning.}
\newblock  (\bibinfo{year}{2002}).
\newblock


\bibitem[Carney and Levin(2003)]%
        {carney2003promoting}
\bibfield{author}{\bibinfo{person}{Russell~N Carney} {and} \bibinfo{person}{Joel~R Levin}.} \bibinfo{year}{2003}\natexlab{}.
\newblock \showarticletitle{Promoting higher-order learning benefits by building lower-order mnemonic connections}.
\newblock \bibinfo{journal}{\emph{Applied Cognitive Psychology: The Official Journal of the Society for Applied Research in Memory and Cognition}} \bibinfo{volume}{17}, \bibinfo{number}{5} (\bibinfo{year}{2003}), \bibinfo{pages}{563--575}.
\newblock


\bibitem[Chen et~al\mbox{.}(2024)]%
        {chen2024learning}
\bibfield{author}{\bibinfo{person}{John Chen}, \bibinfo{person}{Xi Lu}, \bibinfo{person}{Yuzhou Du}, \bibinfo{person}{Michael Rejtig}, \bibinfo{person}{Ruth Bagley}, \bibinfo{person}{Mike Horn}, {and} \bibinfo{person}{Uri Wilensky}.} \bibinfo{year}{2024}\natexlab{}.
\newblock \showarticletitle{Learning agent-based modeling with LLM companions: Experiences of novices and experts using ChatGPT \& NetLogo chat}. In \bibinfo{booktitle}{\emph{Proceedings of the CHI Conference on Human Factors in Computing Systems}}. \bibinfo{pages}{1--18}.
\newblock


\bibitem[Cirigliano(2013)]%
        {cirigliano2013musical}
\bibfield{author}{\bibinfo{person}{Matthew~M Cirigliano}.} \bibinfo{year}{2013}\natexlab{}.
\newblock \showarticletitle{Musical mnemonics in health science: A first look}.
\newblock \bibinfo{journal}{\emph{Medical Teacher}} \bibinfo{volume}{35}, \bibinfo{number}{3} (\bibinfo{year}{2013}), \bibinfo{pages}{e1020--e1026}.
\newblock


\bibitem[Cools et~al\mbox{.}(2023)]%
        {cools2023effects}
\bibfield{author}{\bibinfo{person}{Wouter Cools}, \bibinfo{person}{Ulrike De~Fr{\`e}re}, {and} \bibinfo{person}{Ariane Caplin}.} \bibinfo{year}{2023}\natexlab{}.
\newblock \showarticletitle{Effects of Playing Music during PE on Intrinsic Motivation of Students}.
\newblock \bibinfo{journal}{\emph{Youth}} \bibinfo{volume}{3}, \bibinfo{number}{2} (\bibinfo{year}{2023}), \bibinfo{pages}{654--670}.
\newblock


\bibitem[Das(2023)]%
        {das2023musical}
\bibfield{author}{\bibinfo{person}{Sumantra~S Das}.} \bibinfo{year}{2023}\natexlab{}.
\newblock \showarticletitle{Musical Appeal and Advertising: A Study of Audience Recall and Effectiveness}.
\newblock \bibinfo{journal}{\emph{MediaSpace: DME Media Journal of Communication}} \bibinfo{volume}{4}, \bibinfo{number}{01} (\bibinfo{year}{2023}), \bibinfo{pages}{32--37}.
\newblock


\bibitem[DeLashmutt(2007)]%
        {delashmutt2007study}
\bibfield{author}{\bibinfo{person}{Kathy DeLashmutt}.} \bibinfo{year}{2007}\natexlab{}.
\newblock \showarticletitle{A study of the role of mnemonics in learning mathematics}.
\newblock  (\bibinfo{year}{2007}).
\newblock


\bibitem[Dickey et~al\mbox{.}(2023)]%
        {dickey2023innovating}
\bibfield{author}{\bibinfo{person}{Ethan Dickey}, \bibinfo{person}{Andres Bejarano}, {and} \bibinfo{person}{Chirayu Garg}.} \bibinfo{year}{2023}\natexlab{}.
\newblock \showarticletitle{Innovating Computer Programming Pedagogy: The AI-Lab Framework for Generative AI Adoption}.
\newblock \bibinfo{journal}{\emph{arXiv preprint arXiv:2308.12258}} (\bibinfo{year}{2023}).
\newblock


\bibitem[Dresler et~al\mbox{.}(2017)]%
        {dresler2017mnemonic}
\bibfield{author}{\bibinfo{person}{Martin Dresler}, \bibinfo{person}{William~R Shirer}, \bibinfo{person}{Boris~N Konrad}, \bibinfo{person}{Nils~CJ M{\"u}ller}, \bibinfo{person}{Isabella~C Wagner}, \bibinfo{person}{Guill{\'e}n Fern{\'a}ndez}, \bibinfo{person}{Michael Czisch}, {and} \bibinfo{person}{Michael~D Greicius}.} \bibinfo{year}{2017}\natexlab{}.
\newblock \showarticletitle{Mnemonic training reshapes brain networks to support superior memory}.
\newblock \bibinfo{journal}{\emph{Neuron}} \bibinfo{volume}{93}, \bibinfo{number}{5} (\bibinfo{year}{2017}), \bibinfo{pages}{1227--1235}.
\newblock


\bibitem[Dunlap and Lowenthal(2010)]%
        {joanna2010hot}
\bibfield{author}{\bibinfo{person}{Joanna~C. Dunlap} {and} \bibinfo{person}{Patrick~R. Lowenthal}.} \bibinfo{year}{2010}\natexlab{}.
\newblock \bibinfo{title}{Hot for teacher: Using digital music to enhance students’ experience in online courses}.
\newblock
\newblock


\bibitem[Giannakos et~al\mbox{.}(2024)]%
        {giannakos2024promise}
\bibfield{author}{\bibinfo{person}{Michail Giannakos}, \bibinfo{person}{Roger Azevedo}, \bibinfo{person}{Peter Brusilovsky}, \bibinfo{person}{Mutlu Cukurova}, \bibinfo{person}{Yannis Dimitriadis}, \bibinfo{person}{Davinia Hernandez-Leo}, \bibinfo{person}{Sanna J{\"a}rvel{\"a}}, \bibinfo{person}{Manolis Mavrikis}, {and} \bibinfo{person}{Bart Rienties}.} \bibinfo{year}{2024}\natexlab{}.
\newblock \showarticletitle{The promise and challenges of generative AI in education}.
\newblock \bibinfo{journal}{\emph{Behaviour \& Information Technology}} (\bibinfo{year}{2024}), \bibinfo{pages}{1--27}.
\newblock


\bibitem[Giray(2023)]%
        {giray2023prompt}
\bibfield{author}{\bibinfo{person}{Louie Giray}.} \bibinfo{year}{2023}\natexlab{}.
\newblock \showarticletitle{Prompt engineering with ChatGPT: a guide for academic writers}.
\newblock \bibinfo{journal}{\emph{Annals of biomedical engineering}} \bibinfo{volume}{51}, \bibinfo{number}{12} (\bibinfo{year}{2023}), \bibinfo{pages}{2629--2633}.
\newblock


\bibitem[Groussard et~al\mbox{.}(2010)]%
        {groussard2010music}
\bibfield{author}{\bibinfo{person}{Mathilde Groussard}, \bibinfo{person}{Renaud La~Joie}, \bibinfo{person}{Geraldine Rauchs}, \bibinfo{person}{Brigitte Landeau}, \bibinfo{person}{Gael Chetelat}, \bibinfo{person}{Fausto Viader}, \bibinfo{person}{Beatrice Desgranges}, \bibinfo{person}{Francis Eustache}, {and} \bibinfo{person}{Herve Platel}.} \bibinfo{year}{2010}\natexlab{}.
\newblock \showarticletitle{When music and long-term memory interact: effects of musical expertise on functional and structural plasticity in the hippocampus}.
\newblock \bibinfo{journal}{\emph{PloS one}} \bibinfo{volume}{5}, \bibinfo{number}{10} (\bibinfo{year}{2010}), \bibinfo{pages}{e13225}.
\newblock


\bibitem[Hallam and Rogers(2016)]%
        {hallam2016impact}
\bibfield{author}{\bibinfo{person}{Susan Hallam} {and} \bibinfo{person}{Kevin Rogers}.} \bibinfo{year}{2016}\natexlab{}.
\newblock \showarticletitle{The impact of instrumental music learning on attainment at age 16: A pilot study}.
\newblock \bibinfo{journal}{\emph{British Journal of Music Education}} \bibinfo{volume}{33}, \bibinfo{number}{3} (\bibinfo{year}{2016}), \bibinfo{pages}{247--261}.
\newblock


\bibitem[Halpern and Bartlett(2011)]%
        {halpern2011persistence}
\bibfield{author}{\bibinfo{person}{Andrea~R Halpern} {and} \bibinfo{person}{James~C Bartlett}.} \bibinfo{year}{2011}\natexlab{}.
\newblock \showarticletitle{The persistence of musical memories: A descriptive study of earworms}.
\newblock \bibinfo{journal}{\emph{Music perception}} \bibinfo{volume}{28}, \bibinfo{number}{4} (\bibinfo{year}{2011}), \bibinfo{pages}{425--432}.
\newblock


\bibitem[Hein(1991)]%
        {hein1991constructivist}
\bibfield{author}{\bibinfo{person}{George~E Hein}.} \bibinfo{year}{1991}\natexlab{}.
\newblock \showarticletitle{Constructivist learning theory}.
\newblock \bibinfo{journal}{\emph{Institute for Inquiry. Available at:/http://www. exploratorium. edu/ifi/resources/constructivistlearning. htmlS}} (\bibinfo{year}{1991}).
\newblock


\bibitem[Hodges and O’Connell(2005)]%
        {hodges2005impact}
\bibfield{author}{\bibinfo{person}{Donald~A Hodges} {and} \bibinfo{person}{Debra~S O’Connell}.} \bibinfo{year}{2005}\natexlab{}.
\newblock \showarticletitle{The impact of music education on academic achievement}.
\newblock \bibinfo{journal}{\emph{The University of North Carolina at Greensboro. Retrieved August}}  \bibinfo{volume}{20} (\bibinfo{year}{2005}), \bibinfo{pages}{2010}.
\newblock


\bibitem[Hu et~al\mbox{.}(2021)]%
        {hu2021university}
\bibfield{author}{\bibinfo{person}{Xiao Hu}, \bibinfo{person}{Jing Chen}, {and} \bibinfo{person}{Yuhao Wang}.} \bibinfo{year}{2021}\natexlab{}.
\newblock \showarticletitle{University students’ use of music for learning and well-being: A qualitative study and design implications}.
\newblock \bibinfo{journal}{\emph{Information Processing \& Management}} \bibinfo{volume}{58}, \bibinfo{number}{1} (\bibinfo{year}{2021}), \bibinfo{pages}{102409}.
\newblock


\bibitem[Ikei et~al\mbox{.}(2007)]%
        {ikei2007spatial}
\bibfield{author}{\bibinfo{person}{Yasushi Ikei}, \bibinfo{person}{Hirofumi Ota}, {and} \bibinfo{person}{Takuro Kayahara}.} \bibinfo{year}{2007}\natexlab{}.
\newblock \showarticletitle{Spatial electronic mnemonics: A virtual memory interface}. In \bibinfo{booktitle}{\emph{Human Interface and the Management of Information. Interacting in Information Environments: Symposium on Human Interface 2007, Held as Part of HCI International 2007, Beijing, China, July 22-27, 2007, Proceedings, Part II}}. Springer, \bibinfo{pages}{30--37}.
\newblock


\bibitem[Jordan et~al\mbox{.}(1996)]%
        {jordan1996usability}
\bibfield{author}{\bibinfo{person}{Patrick~W Jordan}, \bibinfo{person}{Bruce Thomas}, \bibinfo{person}{Ian~Lyall McClelland}, {and} \bibinfo{person}{Bernard Weerdmeester}.} \bibinfo{year}{1996}\natexlab{}.
\newblock \bibinfo{booktitle}{\emph{Usability evaluation in industry}}.
\newblock \bibinfo{publisher}{CRC Press}.
\newblock


\bibitem[Kaschel et~al\mbox{.}(2002)]%
        {kaschel2002imagery}
\bibfield{author}{\bibinfo{person}{Reiner Kaschel}, \bibinfo{person}{Sergio~Della Sala}, \bibinfo{person}{Anna Cantagallo}, \bibinfo{person}{Andrea Fahlb{\"o}ck}, \bibinfo{person}{Ritva Laaksonen}, {and} \bibinfo{person}{Miguel Kazen}.} \bibinfo{year}{2002}\natexlab{}.
\newblock \showarticletitle{Imagery mnemonics for the rehabilitation of memory: A randomised group controlled trial}.
\newblock \bibinfo{journal}{\emph{Neuropsychological rehabilitation}} \bibinfo{volume}{12}, \bibinfo{number}{2} (\bibinfo{year}{2002}), \bibinfo{pages}{127--153}.
\newblock


\bibitem[Kaur(2022)]%
        {kaur2022effect}
\bibfield{author}{\bibinfo{person}{Prabhjot Kaur}.} \bibinfo{year}{2022}\natexlab{}.
\newblock \showarticletitle{Effect of mnemonics learning technique on memory: an experimental study on bachelorette nursing students}.
\newblock \bibinfo{journal}{\emph{International Journal of Advanced Education and Research}}  \bibinfo{volume}{11} (\bibinfo{year}{2022}), \bibinfo{pages}{21--26}.
\newblock


\bibitem[Kazemitabaar et~al\mbox{.}(2023)]%
        {kazemitabaar2023studying}
\bibfield{author}{\bibinfo{person}{Majeed Kazemitabaar}, \bibinfo{person}{Justin Chow}, \bibinfo{person}{Carl Ka~To Ma}, \bibinfo{person}{Barbara~J Ericson}, \bibinfo{person}{David Weintrop}, {and} \bibinfo{person}{Tovi Grossman}.} \bibinfo{year}{2023}\natexlab{}.
\newblock \showarticletitle{Studying the effect of AI code generators on supporting novice learners in introductory programming}. In \bibinfo{booktitle}{\emph{Proceedings of the 2023 CHI Conference on Human Factors in Computing Systems}}. \bibinfo{pages}{1--23}.
\newblock


\bibitem[Kelleher et~al\mbox{.}(2007)]%
        {kelleher2007storytelling}
\bibfield{author}{\bibinfo{person}{Caitlin Kelleher}, \bibinfo{person}{Randy Pausch}, {and} \bibinfo{person}{Sara Kiesler}.} \bibinfo{year}{2007}\natexlab{}.
\newblock \showarticletitle{Storytelling alice motivates middle school girls to learn computer programming}. In \bibinfo{booktitle}{\emph{Proceedings of the SIGCHI conference on Human factors in computing systems}}. \bibinfo{pages}{1455--1464}.
\newblock


\bibitem[King-Sears et~al\mbox{.}(1992)]%
        {king1992toward}
\bibfield{author}{\bibinfo{person}{Margaret~E King-Sears}, \bibinfo{person}{Cecil~D Mercer}, {and} \bibinfo{person}{Paul~T Sindelar}.} \bibinfo{year}{1992}\natexlab{}.
\newblock \showarticletitle{Toward independence with keyword mnemonics: A strategy for science vocabulary instruction}.
\newblock \bibinfo{journal}{\emph{Remedial and Special Education}} \bibinfo{volume}{13}, \bibinfo{number}{5} (\bibinfo{year}{1992}), \bibinfo{pages}{22--33}.
\newblock


\bibitem[Koksal et~al\mbox{.}(2013)]%
        {koksal2013impact}
\bibfield{author}{\bibinfo{person}{Onur Koksal}, \bibinfo{person}{Ali~Murat Sunbul}, \bibinfo{person}{Yunus~Emre Ozturk}, {and} \bibinfo{person}{Musa Ozata}.} \bibinfo{year}{2013}\natexlab{}.
\newblock \showarticletitle{The Impact of Mnemonic Devices on Attainment and Recall in Basic Knowledge Acquisition in Nursing Education.}
\newblock \bibinfo{journal}{\emph{Mevlana International Journal of Education}} \bibinfo{volume}{3}, \bibinfo{number}{4} (\bibinfo{year}{2013}).
\newblock


\bibitem[Kuo et~al\mbox{.}(2006)]%
        {kuo2006human}
\bibfield{author}{\bibinfo{person}{Cynthia Kuo}, \bibinfo{person}{Sasha Romanosky}, {and} \bibinfo{person}{Lorrie~Faith Cranor}.} \bibinfo{year}{2006}\natexlab{}.
\newblock \showarticletitle{Human selection of mnemonic phrase-based passwords}. In \bibinfo{booktitle}{\emph{Proceedings of the second symposium on Usable privacy and security}}. \bibinfo{pages}{67--78}.
\newblock


\bibitem[Levin(1993)]%
        {levin1993mnemonic}
\bibfield{author}{\bibinfo{person}{Joel~R Levin}.} \bibinfo{year}{1993}\natexlab{}.
\newblock \showarticletitle{Mnemonic strategies and classroom learning: A twenty-year report card}.
\newblock \bibinfo{journal}{\emph{The Elementary School Journal}} \bibinfo{volume}{94}, \bibinfo{number}{2} (\bibinfo{year}{1993}), \bibinfo{pages}{235--244}.
\newblock


\bibitem[Lewandowsky et~al\mbox{.}(2012)]%
        {lewandowsky2012misinformation}
\bibfield{author}{\bibinfo{person}{Stephan Lewandowsky}, \bibinfo{person}{Ullrich~KH Ecker}, \bibinfo{person}{Colleen~M Seifert}, \bibinfo{person}{Norbert Schwarz}, {and} \bibinfo{person}{John Cook}.} \bibinfo{year}{2012}\natexlab{}.
\newblock \showarticletitle{Misinformation and its correction: Continued influence and successful debiasing}.
\newblock \bibinfo{journal}{\emph{Psychological science in the public interest}} \bibinfo{volume}{13}, \bibinfo{number}{3} (\bibinfo{year}{2012}), \bibinfo{pages}{106--131}.
\newblock


\bibitem[Lewis(2018)]%
        {lewis2018system}
\bibfield{author}{\bibinfo{person}{James~R Lewis}.} \bibinfo{year}{2018}\natexlab{}.
\newblock \showarticletitle{The system usability scale: past, present, and future}.
\newblock \bibinfo{journal}{\emph{International Journal of Human--Computer Interaction}} \bibinfo{volume}{34}, \bibinfo{number}{7} (\bibinfo{year}{2018}), \bibinfo{pages}{577--590}.
\newblock


\bibitem[Lewis~Jr et~al\mbox{.}(2018)]%
        {lewis2018importance}
\bibfield{author}{\bibinfo{person}{James~B Lewis~Jr}, \bibinfo{person}{Rebekah Mulligan}, {and} \bibinfo{person}{Neal Kraus}.} \bibinfo{year}{2018}\natexlab{}.
\newblock \showarticletitle{The importance of medical mnemonics in medicine}.
\newblock \bibinfo{journal}{\emph{Pharos}}  \bibinfo{volume}{2018} (\bibinfo{year}{2018}), \bibinfo{pages}{30--35}.
\newblock


\bibitem[Li et~al\mbox{.}(2024)]%
        {li2024dawn}
\bibfield{author}{\bibinfo{person}{Junyi Li}, \bibinfo{person}{Jie Chen}, \bibinfo{person}{Ruiyang Ren}, \bibinfo{person}{Xiaoxue Cheng}, \bibinfo{person}{Wayne~Xin Zhao}, \bibinfo{person}{Jian-Yun Nie}, {and} \bibinfo{person}{Ji-Rong Wen}.} \bibinfo{year}{2024}\natexlab{}.
\newblock \showarticletitle{The dawn after the dark: An empirical study on factuality hallucination in large language models}.
\newblock \bibinfo{journal}{\emph{arXiv preprint arXiv:2401.03205}} (\bibinfo{year}{2024}).
\newblock


\bibitem[Manalo et~al\mbox{.}(2004)]%
        {manalo2004using}
\bibfield{author}{\bibinfo{person}{Emmanuel Manalo}, \bibinfo{person}{Satomi Mizutani}, {and} \bibinfo{person}{Julie Trafford}.} \bibinfo{year}{2004}\natexlab{}.
\newblock \showarticletitle{Using mnemonics to facilitate learning of Japanese script characters}.
\newblock \bibinfo{journal}{\emph{JALT Journal}} \bibinfo{volume}{26}, \bibinfo{number}{1} (\bibinfo{year}{2004}), \bibinfo{pages}{55--77}.
\newblock


\bibitem[Margulieux et~al\mbox{.}(2024)]%
        {margulieux2024recommendations}
\bibfield{author}{\bibinfo{person}{Lauren~E Margulieux}, \bibinfo{person}{Ben~Rydal Shapiro}, \bibinfo{person}{Brendan~D Calandra}, {et~al\mbox{.}}} \bibinfo{year}{2024}\natexlab{}.
\newblock \showarticletitle{Recommendations for Computer Science Education in Colleges of Education}.
\newblock \bibinfo{journal}{\emph{Authorea Preprints}} (\bibinfo{year}{2024}).
\newblock


\bibitem[Marvin et~al\mbox{.}(2023)]%
        {marvin2023prompt}
\bibfield{author}{\bibinfo{person}{Ggaliwango Marvin}, \bibinfo{person}{Nakayiza Hellen}, \bibinfo{person}{Daudi Jjingo}, {and} \bibinfo{person}{Joyce Nakatumba-Nabende}.} \bibinfo{year}{2023}\natexlab{}.
\newblock \showarticletitle{Prompt engineering in large language models}. In \bibinfo{booktitle}{\emph{International conference on data intelligence and cognitive informatics}}. Springer, \bibinfo{pages}{387--402}.
\newblock


\bibitem[Mash and Edwards(2020)]%
        {mash2020creating}
\bibfield{author}{\bibinfo{person}{Bob Mash} {and} \bibinfo{person}{Jill Edwards}.} \bibinfo{year}{2020}\natexlab{}.
\newblock \showarticletitle{Creating a learning environment in your practice or facility}.
\newblock \bibinfo{journal}{\emph{South African Family Practice}} \bibinfo{volume}{62}, \bibinfo{number}{3} (\bibinfo{year}{2020}).
\newblock


\bibitem[McCabe(2015)]%
        {mccabe2015learning}
\bibfield{author}{\bibinfo{person}{Jennifer~A McCabe}.} \bibinfo{year}{2015}\natexlab{}.
\newblock \showarticletitle{Learning the brain in introductory psychology: Examining the generation effect for mnemonics and examples}.
\newblock \bibinfo{journal}{\emph{Teaching of Psychology}} \bibinfo{volume}{42}, \bibinfo{number}{3} (\bibinfo{year}{2015}), \bibinfo{pages}{203--210}.
\newblock


\bibitem[McCabe et~al\mbox{.}(2013)]%
        {mccabe2013psychology}
\bibfield{author}{\bibinfo{person}{Jennifer~A McCabe}, \bibinfo{person}{Kelsey~L Osha}, \bibinfo{person}{Jennifer~A Roche}, {and} \bibinfo{person}{Jonathan~A Susser}.} \bibinfo{year}{2013}\natexlab{}.
\newblock \showarticletitle{Psychology students’ knowledge and use of mnemonics}.
\newblock \bibinfo{journal}{\emph{Teaching of psychology}} \bibinfo{volume}{40}, \bibinfo{number}{3} (\bibinfo{year}{2013}), \bibinfo{pages}{183--192}.
\newblock


\bibitem[Mega et~al\mbox{.}(2014)]%
        {mega2014makes}
\bibfield{author}{\bibinfo{person}{Carolina Mega}, \bibinfo{person}{Lucia Ronconi}, {and} \bibinfo{person}{Rossana De~Beni}.} \bibinfo{year}{2014}\natexlab{}.
\newblock \showarticletitle{What makes a good student? How emotions, self-regulated learning, and motivation contribute to academic achievement.}
\newblock \bibinfo{journal}{\emph{Journal of educational psychology}} \bibinfo{volume}{106}, \bibinfo{number}{1} (\bibinfo{year}{2014}), \bibinfo{pages}{121}.
\newblock


\bibitem[Nie et~al\mbox{.}(2023)]%
        {nie2023research}
\bibfield{author}{\bibinfo{person}{Junting Nie} {et~al\mbox{.}}} \bibinfo{year}{2023}\natexlab{}.
\newblock \showarticletitle{Research on the relationship between achievement motivation and individual emotional state: the promoting effect of positive emotions}.
\newblock \bibinfo{journal}{\emph{Applied \& Educational Psychology}} \bibinfo{volume}{4}, \bibinfo{number}{10} (\bibinfo{year}{2023}), \bibinfo{pages}{94--100}.
\newblock


\bibitem[Perkins et~al\mbox{.}(1992)]%
        {perkins1992transfer}
\bibfield{author}{\bibinfo{person}{David~N Perkins}, \bibinfo{person}{Gavriel Salomon}, {et~al\mbox{.}}} \bibinfo{year}{1992}\natexlab{}.
\newblock \showarticletitle{Transfer of learning}.
\newblock \bibinfo{journal}{\emph{International encyclopedia of education}}  \bibinfo{volume}{2} (\bibinfo{year}{1992}), \bibinfo{pages}{6452--6457}.
\newblock


\bibitem[Perrault et~al\mbox{.}(2015)]%
        {perrault2015physical}
\bibfield{author}{\bibinfo{person}{Simon~T Perrault}, \bibinfo{person}{Eric Lecolinet}, \bibinfo{person}{Yoann~Pascal Bourse}, \bibinfo{person}{Shengdong Zhao}, {and} \bibinfo{person}{Yves Guiard}.} \bibinfo{year}{2015}\natexlab{}.
\newblock \showarticletitle{Physical loci: Leveraging spatial, object and semantic memory for command selection}. In \bibinfo{booktitle}{\emph{Proceedings of the 33rd annual acm conference on human factors in computing systems}}. \bibinfo{pages}{299--308}.
\newblock


\bibitem[Putnam(2015)]%
        {putnam2015mnemonics}
\bibfield{author}{\bibinfo{person}{Adam~L Putnam}.} \bibinfo{year}{2015}\natexlab{}.
\newblock \showarticletitle{Mnemonics in education: Current research and applications.}
\newblock \bibinfo{journal}{\emph{Translational Issues in Psychological Science}} \bibinfo{volume}{1}, \bibinfo{number}{2} (\bibinfo{year}{2015}), \bibinfo{pages}{130}.
\newblock


\bibitem[Qi(2023)]%
        {qi2023role}
\bibfield{author}{\bibinfo{person}{Jing Qi}.} \bibinfo{year}{2023}\natexlab{}.
\newblock \showarticletitle{The Role of Chinese Music in Shaping Students’ Creative Thinking}.
\newblock \bibinfo{journal}{\emph{Educational Process: International Journal (EDUPIJ)}} \bibinfo{volume}{12}, \bibinfo{number}{2} (\bibinfo{year}{2023}), \bibinfo{pages}{111--123}.
\newblock


\bibitem[Raugh and Atkinson(1975)]%
        {raugh1975mnemonic}
\bibfield{author}{\bibinfo{person}{Michael~R Raugh} {and} \bibinfo{person}{Richard~C Atkinson}.} \bibinfo{year}{1975}\natexlab{}.
\newblock \showarticletitle{A mnemonic method for learning a second-language vocabulary.}
\newblock \bibinfo{journal}{\emph{Journal of Educational Psychology}} \bibinfo{volume}{67}, \bibinfo{number}{1} (\bibinfo{year}{1975}), \bibinfo{pages}{1}.
\newblock


\bibitem[Renninger et~al\mbox{.}(2014)]%
        {renninger2014role}
\bibfield{author}{\bibinfo{person}{K~Ann Renninger}, \bibinfo{person}{Suzanne Hidi}, \bibinfo{person}{Andreas Krapp}, {and} \bibinfo{person}{Ann Renninger}.} \bibinfo{year}{2014}\natexlab{}.
\newblock \bibinfo{booktitle}{\emph{The role of interest in learning and development}}.
\newblock \bibinfo{publisher}{Psychology Press}.
\newblock


\bibitem[Richardson(1995)]%
        {richardson1995efficacy}
\bibfield{author}{\bibinfo{person}{John~TE Richardson}.} \bibinfo{year}{1995}\natexlab{}.
\newblock \showarticletitle{The efficacy of imagery mnemonics in memory remediation}.
\newblock \bibinfo{journal}{\emph{Neuropsychologia}} \bibinfo{volume}{33}, \bibinfo{number}{11} (\bibinfo{year}{1995}), \bibinfo{pages}{1345--1357}.
\newblock


\bibitem[Roden et~al\mbox{.}(2012)]%
        {roden2012effects}
\bibfield{author}{\bibinfo{person}{Ingo Roden}, \bibinfo{person}{Gunter Kreutz}, {and} \bibinfo{person}{Stephan Bongard}.} \bibinfo{year}{2012}\natexlab{}.
\newblock \showarticletitle{Effects of a school-based instrumental music program on verbal and visual memory in primary school children: a longitudinal study}.
\newblock \bibinfo{journal}{\emph{Frontiers in psychology}}  \bibinfo{volume}{3} (\bibinfo{year}{2012}), \bibinfo{pages}{572}.
\newblock


\bibitem[Roediger(1980)]%
        {roediger1980effectiveness}
\bibfield{author}{\bibinfo{person}{Henry~L Roediger}.} \bibinfo{year}{1980}\natexlab{}.
\newblock \showarticletitle{The effectiveness of four mnemonics in ordering recall.}
\newblock \bibinfo{journal}{\emph{Journal of Experimental Psychology: Human Learning and Memory}} \bibinfo{volume}{6}, \bibinfo{number}{5} (\bibinfo{year}{1980}), \bibinfo{pages}{558}.
\newblock


\bibitem[Romeike(2019)]%
        {romeike2019role}
\bibfield{author}{\bibinfo{person}{Ralf Romeike}.} \bibinfo{year}{2019}\natexlab{}.
\newblock \showarticletitle{The role of computer science education for understanding and shaping the digital society}. In \bibinfo{booktitle}{\emph{Sustainable ICT, Education and Learning: IFIP WG 3.4 International Conference, SUZA 2019, Zanzibar, Tanzania, April 25--27, 2019, Revised Selected Papers 1}}. Springer, \bibinfo{pages}{167--176}.
\newblock


\bibitem[Sahakian et~al\mbox{.}(1990)]%
        {sahakian1990sparing}
\bibfield{author}{\bibinfo{person}{BJ Sahakian}, \bibinfo{person}{JJ Downes}, \bibinfo{person}{S Eagger}, \bibinfo{person}{JL Everden}, \bibinfo{person}{R Levy}, \bibinfo{person}{MP Philpot}, \bibinfo{person}{AC Roberts}, {and} \bibinfo{person}{TW Robbins}.} \bibinfo{year}{1990}\natexlab{}.
\newblock \showarticletitle{Sparing of attentional relative to mnemonic function in a subgroup of patients with dementia of the Alzheimer type}.
\newblock \bibinfo{journal}{\emph{Neuropsychologia}} \bibinfo{volume}{28}, \bibinfo{number}{11} (\bibinfo{year}{1990}), \bibinfo{pages}{1197--1213}.
\newblock


\bibitem[Scripp(2002)]%
        {scripp2002overview}
\bibfield{author}{\bibinfo{person}{Larry Scripp}.} \bibinfo{year}{2002}\natexlab{}.
\newblock \showarticletitle{An overview of research on music and learning}.
\newblock \bibinfo{journal}{\emph{Critical links: Learning in the arts and student academic and social development}} (\bibinfo{year}{2002}), \bibinfo{pages}{132--136}.
\newblock


\bibitem[Shackelford and LeBlanc(1997)]%
        {shackelford1997introducing}
\bibfield{author}{\bibinfo{person}{Russell~L Shackelford} {and} \bibinfo{person}{RJ LeBlanc}.} \bibinfo{year}{1997}\natexlab{}.
\newblock \showarticletitle{Introducing computer science fundamentals before programming}. In \bibinfo{booktitle}{\emph{Proceedings frontiers in education 1997 27th annual conference. teaching and learning in an era of change}}, Vol.~\bibinfo{volume}{1}. IEEE, \bibinfo{pages}{285--289}.
\newblock


\bibitem[Shams and Seitz(2008)]%
        {shams2008benefits}
\bibfield{author}{\bibinfo{person}{Ladan Shams} {and} \bibinfo{person}{Aaron~R Seitz}.} \bibinfo{year}{2008}\natexlab{}.
\newblock \showarticletitle{Benefits of multisensory learning}.
\newblock \bibinfo{journal}{\emph{Trends in cognitive sciences}} \bibinfo{volume}{12}, \bibinfo{number}{11} (\bibinfo{year}{2008}), \bibinfo{pages}{411--417}.
\newblock


\bibitem[Shastri(2020)]%
        {shastri2020machine}
\bibfield{author}{\bibinfo{person}{Dvijesh~J Shastri}.} \bibinfo{year}{2020}\natexlab{}.
\newblock \showarticletitle{Machine learning for non-programmers}. In \bibinfo{booktitle}{\emph{Extended Abstracts of the 2020 CHI Conference on Human Factors in Computing Systems}}. \bibinfo{pages}{1--3}.
\newblock


\bibitem[Shuang-yi(2018)]%
        {shuang2018forced}
\bibfield{author}{\bibinfo{person}{C Shuang-yi}.} \bibinfo{year}{2018}\natexlab{}.
\newblock \showarticletitle{Forced Learning: Manifestations, Hazards, and Coping Strategies}.
\newblock \bibinfo{journal}{\emph{US-China Education Review B}} \bibinfo{volume}{8}, \bibinfo{number}{9} (\bibinfo{year}{2018}), \bibinfo{pages}{404--411}.
\newblock


\bibitem[Simon(1978)]%
        {simon1978information}
\bibfield{author}{\bibinfo{person}{Herbert~A Simon}.} \bibinfo{year}{1978}\natexlab{}.
\newblock \showarticletitle{Information-processing theory of human problem solving}.
\newblock \bibinfo{journal}{\emph{Handbook of learning and cognitive processes}}  \bibinfo{volume}{5} (\bibinfo{year}{1978}), \bibinfo{pages}{271--295}.
\newblock


\bibitem[Slamecka and Graf(1978)]%
        {slamecka1978generation}
\bibfield{author}{\bibinfo{person}{Norman~J Slamecka} {and} \bibinfo{person}{Peter Graf}.} \bibinfo{year}{1978}\natexlab{}.
\newblock \showarticletitle{The generation effect: Delineation of a phenomenon.}
\newblock \bibinfo{journal}{\emph{Journal of experimental Psychology: Human learning and Memory}} \bibinfo{volume}{4}, \bibinfo{number}{6} (\bibinfo{year}{1978}), \bibinfo{pages}{592}.
\newblock


\bibitem[Smith and Phillips~Jr(2001)]%
        {smith2001age}
\bibfield{author}{\bibinfo{person}{Malcolm~C Smith} {and} \bibinfo{person}{Mark~R Phillips~Jr}.} \bibinfo{year}{2001}\natexlab{}.
\newblock \showarticletitle{Age differences in memory for radio advertisements: the role of mnemonics}.
\newblock \bibinfo{journal}{\emph{Journal of Business Research}} \bibinfo{volume}{53}, \bibinfo{number}{2} (\bibinfo{year}{2001}), \bibinfo{pages}{103--109}.
\newblock


\bibitem[Soll and Kobras(2022)]%
        {soll2022were}
\bibfield{author}{\bibinfo{person}{Marcus Soll} {and} \bibinfo{person}{Louis Kobras}.} \bibinfo{year}{2022}\natexlab{}.
\newblock \showarticletitle{What Were We Expecting? Analysing Expectations of German University Teachers of Study Beginners in Computer Science as Experienced by Students}. In \bibinfo{booktitle}{\emph{2022 IEEE German Education Conference (GeCon)}}. IEEE, \bibinfo{pages}{1--6}.
\newblock


\bibitem[Su et~al\mbox{.}(2012)]%
        {su2012linear}
\bibfield{author}{\bibinfo{person}{Xiaogang Su}, \bibinfo{person}{Xin Yan}, {and} \bibinfo{person}{Chih-Ling Tsai}.} \bibinfo{year}{2012}\natexlab{}.
\newblock \showarticletitle{Linear regression}.
\newblock \bibinfo{journal}{\emph{Wiley Interdisciplinary Reviews: Computational Statistics}} \bibinfo{volume}{4}, \bibinfo{number}{3} (\bibinfo{year}{2012}), \bibinfo{pages}{275--294}.
\newblock


\bibitem[Tang et~al\mbox{.}(2024)]%
        {tang2024vizgroup}
\bibfield{author}{\bibinfo{person}{Xiaohang Tang}, \bibinfo{person}{Sam Wong}, \bibinfo{person}{Kevin Pu}, \bibinfo{person}{Xi Chen}, \bibinfo{person}{Yalong Yang}, {and} \bibinfo{person}{Yan Chen}.} \bibinfo{year}{2024}\natexlab{}.
\newblock \showarticletitle{VizGroup: An AI-Assisted Event-Driven System for Real-Time Collaborative Programming Learning Analytics}.
\newblock \bibinfo{journal}{\emph{arXiv preprint arXiv:2404.08743}} (\bibinfo{year}{2024}).
\newblock


\bibitem[Tervaniemi et~al\mbox{.}(2018)]%
        {tervaniemi2018promises}
\bibfield{author}{\bibinfo{person}{Mari Tervaniemi}, \bibinfo{person}{Sha Tao}, {and} \bibinfo{person}{Minna Huotilainen}.} \bibinfo{year}{2018}\natexlab{}.
\newblock \showarticletitle{Promises of music in education?}. In \bibinfo{booktitle}{\emph{Frontiers in Education}}, Vol.~\bibinfo{volume}{3}. Frontiers Media SA, \bibinfo{pages}{74}.
\newblock


\bibitem[Thalmann et~al\mbox{.}(2019)]%
        {thalmann2019does}
\bibfield{author}{\bibinfo{person}{Mirko Thalmann}, \bibinfo{person}{Alessandra~S Souza}, {and} \bibinfo{person}{Klaus Oberauer}.} \bibinfo{year}{2019}\natexlab{}.
\newblock \showarticletitle{How does chunking help working memory?}
\newblock \bibinfo{journal}{\emph{Journal of Experimental Psychology: Learning, Memory, and Cognition}} \bibinfo{volume}{45}, \bibinfo{number}{1} (\bibinfo{year}{2019}), \bibinfo{pages}{37}.
\newblock


\bibitem[Tobias and Duffy(2009)]%
        {tobias2009constructivist}
\bibfield{author}{\bibinfo{person}{Sigmund Tobias} {and} \bibinfo{person}{Thomas~M Duffy}.} \bibinfo{year}{2009}\natexlab{}.
\newblock \showarticletitle{Constructivist instruction}.
\newblock \bibinfo{journal}{\emph{Success or failure}} (\bibinfo{year}{2009}).
\newblock


\bibitem[Tucker(2003)]%
        {10.1145/2593247}
\bibfield{author}{\bibinfo{person}{Allen Tucker}.} \bibinfo{year}{2003}\natexlab{}.
\newblock \bibinfo{booktitle}{\emph{A Model Curriculum for K--12 Computer Science: Final Report of the ACM K--12 Task Force Curriculum Committee}}.
\newblock \bibinfo{type}{{T}echnical {R}eport}. \bibinfo{address}{New York, NY, USA}.
\newblock
\showISBNx{1581138377}


\bibitem[Von~Neumann(1945)]%
        {von1945neumann}
\bibfield{author}{\bibinfo{person}{John Von~Neumann}.} \bibinfo{year}{1945}\natexlab{}.
\newblock \showarticletitle{Von neumann architecture}.
\newblock \bibinfo{journal}{\emph{Online http://en. wikipedia. org/wiki/Von\_Neumann\_architecture}}  \bibinfo{volume}{8} (\bibinfo{year}{1945}).
\newblock


\bibitem[Wan et~al\mbox{.}(2020)]%
        {wan2020smileycluster}
\bibfield{author}{\bibinfo{person}{Xiaoyu Wan}, \bibinfo{person}{Xiaofei Zhou}, \bibinfo{person}{Zaiqiao Ye}, \bibinfo{person}{Chase~K Mortensen}, {and} \bibinfo{person}{Zhen Bai}.} \bibinfo{year}{2020}\natexlab{}.
\newblock \showarticletitle{SmileyCluster: supporting accessible machine learning in K-12 scientific discovery}. In \bibinfo{booktitle}{\emph{Proceedings of the interaction design and children conference}}. \bibinfo{pages}{23--35}.
\newblock


\bibitem[Wang et~al\mbox{.}(2021)]%
        {wang2021puzzleme}
\bibfield{author}{\bibinfo{person}{April~Yi Wang}, \bibinfo{person}{Yan Chen}, \bibinfo{person}{John Joon~Young Chung}, \bibinfo{person}{Christopher Brooks}, {and} \bibinfo{person}{Steve Oney}.} \bibinfo{year}{2021}\natexlab{}.
\newblock \showarticletitle{Puzzleme: Leveraging peer assessment for in-class programming exercises}.
\newblock \bibinfo{journal}{\emph{Proceedings of the ACM on Human-Computer Interaction}} \bibinfo{volume}{5}, \bibinfo{number}{CSCW2} (\bibinfo{year}{2021}), \bibinfo{pages}{1--24}.
\newblock


\bibitem[Wang et~al\mbox{.}(2012)]%
        {wang2012block}
\bibfield{author}{\bibinfo{person}{Danli Wang}, \bibinfo{person}{Yang Zhang}, \bibinfo{person}{Tianyuan Gu}, \bibinfo{person}{Liang He}, {and} \bibinfo{person}{Hongan Wang}.} \bibinfo{year}{2012}\natexlab{}.
\newblock \showarticletitle{E-Block: a tangible programming tool for children}. In \bibinfo{booktitle}{\emph{Adjunct proceedings of the 25th annual ACM symposium on User interface software and technology}}. \bibinfo{pages}{71--72}.
\newblock


\bibitem[Wilson(2016)]%
        {wilson2016anderson}
\bibfield{author}{\bibinfo{person}{Leslie~Owen Wilson}.} \bibinfo{year}{2016}\natexlab{}.
\newblock \showarticletitle{Anderson and Krathwohl--Bloom’s taxonomy revised}.
\newblock \bibinfo{journal}{\emph{Understanding the new version of Bloom's taxonomy}} (\bibinfo{year}{2016}).
\newblock


\bibitem[Winkler et~al\mbox{.}(2020)]%
        {winkler2020sara}
\bibfield{author}{\bibinfo{person}{Rainer Winkler}, \bibinfo{person}{Sebastian Hobert}, \bibinfo{person}{Antti Salovaara}, \bibinfo{person}{Matthias S{\"o}llner}, {and} \bibinfo{person}{Jan~Marco Leimeister}.} \bibinfo{year}{2020}\natexlab{}.
\newblock \showarticletitle{Sara, the lecturer: Improving learning in online education with a scaffolding-based conversational agent}. In \bibinfo{booktitle}{\emph{Proceedings of the 2020 CHI conference on human factors in computing systems}}. \bibinfo{pages}{1--14}.
\newblock


\bibitem[Woo and Mirkovic(2016)]%
        {woo2016improving}
\bibfield{author}{\bibinfo{person}{Simon~S Woo} {and} \bibinfo{person}{Jelena Mirkovic}.} \bibinfo{year}{2016}\natexlab{}.
\newblock \showarticletitle{Improving recall and security of passphrases through use of mnemonics}. In \bibinfo{booktitle}{\emph{Proceedings of the 10th International Conference on Passwords (Passwords)}}.
\newblock


\bibitem[Yalch(1991)]%
        {yalch1991memory}
\bibfield{author}{\bibinfo{person}{Richard~F Yalch}.} \bibinfo{year}{1991}\natexlab{}.
\newblock \showarticletitle{Memory in a jingle jungle: Music as a mnemonic device in communicating advertising slogans.}
\newblock \bibinfo{journal}{\emph{Journal of Applied Psychology}} \bibinfo{volume}{76}, \bibinfo{number}{2} (\bibinfo{year}{1991}), \bibinfo{pages}{268}.
\newblock


\bibitem[Yang(2023)]%
        {yang2023impact}
\bibfield{author}{\bibinfo{person}{Jing Yang}.} \bibinfo{year}{2023}\natexlab{}.
\newblock \showarticletitle{The Impact of Music Education on Students’ Academic Performance and Academic Motivation: A Quantitative Study}.
\newblock \bibinfo{journal}{\emph{International Journal of Social Science and Humanities Research}} \bibinfo{volume}{11}, \bibinfo{number}{4} (\bibinfo{year}{2023}), \bibinfo{pages}{121--125}.
\newblock


\bibitem[Yates(1966)]%
        {yates1966art}
\bibfield{author}{\bibinfo{person}{Frances Yates}.} \bibinfo{year}{1966}\natexlab{}.
\newblock \bibinfo{booktitle}{\emph{The Art of Memory}}.
\newblock \bibinfo{publisher}{University of Chicago Press}, \bibinfo{address}{Chicago, IL}.
\newblock


\bibitem[Yeoh(2013)]%
        {yeoh2013musical}
\bibfield{author}{\bibinfo{person}{Miranda~P Yeoh}.} \bibinfo{year}{2013}\natexlab{}.
\newblock \showarticletitle{Musical mnemonics to facilitate learning of matriculation biology: The Calvin Cycle}.
\newblock \bibinfo{journal}{\emph{molecules}}  \bibinfo{volume}{3} (\bibinfo{year}{2013}), \bibinfo{pages}{3C}.
\newblock


\bibitem[Yeoh(2014a)]%
        {yeoh2014musical1}
\bibfield{author}{\bibinfo{person}{Miranda~P Yeoh}.} \bibinfo{year}{2014}\natexlab{a}.
\newblock \showarticletitle{Musical Mnemonics to Facilitate Learning of Protein Synthesis: Translation}.
\newblock \bibinfo{journal}{\emph{Sainsab, Journal of the Association of Science and Mathematics Education Penang}}  \bibinfo{volume}{17} (\bibinfo{year}{2014}), \bibinfo{pages}{1--11}.
\newblock


\bibitem[Yeoh(2014b)]%
        {yeoh2014musical}
\bibfield{author}{\bibinfo{person}{Miranda~P Yeoh}.} \bibinfo{year}{2014}\natexlab{b}.
\newblock \showarticletitle{Musical mnemonics to facilitate learning of transcription of RNA}.
\newblock \bibinfo{journal}{\emph{Learning science and mathematics}}  \bibinfo{volume}{9} (\bibinfo{year}{2014}), \bibinfo{pages}{24--34}.
\newblock


\end{thebibliography}

\appendix
\section{Appendix}

\begin{figure}[H]
    \centering
    \includegraphics[width=0.75\linewidth]{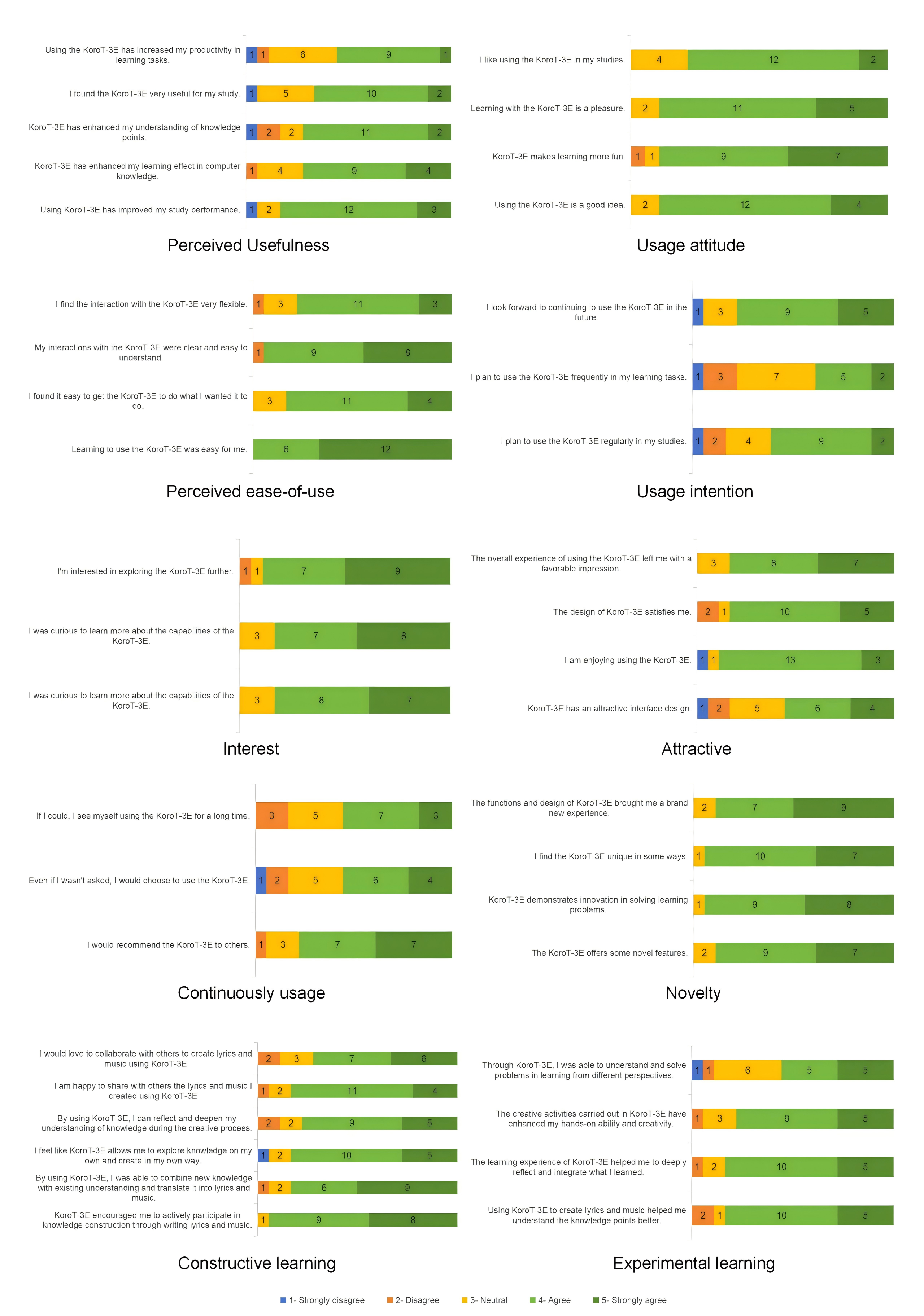}
    \caption{Result of Scale}
    \label{scale}
\end{figure}

\end{document}